\documentclass{aa}  
\usepackage{graphicx,upgreek,txfonts,aas_macros}
\usepackage[dvipsnames]{xcolor}
\usepackage[colorlinks=true,allcolors=blue]{hyperref}
\usepackage{comment}
\usepackage{booktabs}
\begin{document} 

%%%%%%%%%%%%%% MY DEFINITIONS
\def\xmm {\textit{XMM--Newton}}
\def\chandra {\textit{Chandra}}
\def\swift {\textit{Swift}}
\def\frm {\textit{Fermi}}
\def\igr {\textit{INTEGRAL}}
\def\sax {\textit{BeppoSAX}}
\def\xte {\textit{RXTE}}
\def\rst {\textit{ROSAT}}
\def\asca {\textit{ASCA}}
\def\hst {\textit{HST}}
\def\nst {\textit{NuSTAR}}
\def\ero {\textit{eROSITA}}
\def\srcextended {\mbox{CXOGlb\,J002415.8--720436}}
\def\src {W2}
\def\flux {\mbox{erg cm$^{-2}$ s$^{-1}$}}
\def\lum {\mbox{erg s$^{-1}$}}
\def\nh {$N_{\rm H}$}

\newcommand{\rob}[1]{\textcolor{red}{Roberta: #1 }} 
\newcommand{\giallo}[1]{\textcolor{orange}{Giallo: #1 }} 
\newcommand{\matt}[1]{\textcolor{olive}{Matt: #1 }} 
\newcommand{\paolo}[1]{\textcolor{blue}{#1}}
\newcommand{\nicola}[1]{\textcolor{green}{#1}} 
\newcommand{\domitilla}[1]{\textcolor{Green}{#1}} 
%%%%%%%%%%%%%%%%%%%%%%%%%%%%%%%%%%%%%%

\title{On the nature of the 2.4\,hr-period eclipsing cataclysmic variable \src\ in 47\,Tuc}

\titlerunning{On the nature of the 2.4\,hr-period eclipsing cataclysmic variable \src\ in 47\,Tuc}
\authorrunning{R. Amato et al.}

   \author{R.~Amato\inst{1,2} 
   \and N.~La Palombara\inst{3} 
   \and M.~Imbrogno\inst{4,2}
   \and G.L.~Israel\inst{2}
   \and P.~Esposito\inst{5}
   \and D.~de Martino\inst{6}
   \and \\ N.A.~Webb\inst{1}
   \and R.~Iaria\inst{7}
   }

   \institute{IRAP, CNRS, Université de Toulouse, CNES, 9 Avenue du Colonel Roche, 31028 Toulouse, France\\
   e-mail: \href{mailto:roberta.amato@inaf.it}{roberta.amato@inaf.it}
   \and INAF-Osservatorio Astronomico di Roma, via Frascati 33, 00078 Monteporzio Catone, Italy
   \and INAF-Istituto di Astrofisica Spaziale e Fisica Cosmica di Milano, via A. Corti 12, 20133 Milano, Italy
   \and Dipartimento di Fisica, Università degli Studi di Roma ``Tor Vergata'', via della Ricerca Scientifica 1, I-00133 Rome, Italy
   \and Scuola Universitaria Superiore IUSS Pavia, Palazzo del Broletto, piazza della Vittoria 15, 27100 Pavia, Italy
   \and INAF-Osservatorio Astronomico di Capodimonte, salita Moiariello 16, I-80131, Napoli, Italy
   \and Dipartimento di fisica e chimica - Emilio  Segr\`{e}, Universit\`a  degli Studi di Palermo, Via Archirafi 36, 90123, Palermo, Italy}

  \date{Received DD Month YYYY; accepted DD Month YYYY}

  \abstract{\src{} (\srcextended{}) is a cataclysmic variable (CV) in the Galactic globular cluster 47 Tucanae. Its modulation was discovered within the CATS@BAR project. The source shows all the properties of magnetic CVs, but whether it is a polar or an intermediate polar is still a matter of debate.}{This paper investigates the spectral and temporal properties of the source, using all archival X-ray data from \chandra{} and \ero{} Early Data Release, to establish whether the source falls within the category of polars or intermediate polars.}{We fitted \chandra{} archival spectra with three different models: a power law, a bremsstrahlung and an optically thin thermal plasma. We also explored the temporal properties of the source with searches for pulsations with a power spectral density analysis and a Rayleigh test ($Z_n^2$).}
  {\src{} displays a mean luminosity of $\sim10^{32}$ \lum{} over a 20-year span, despite lower values in a few epochs. The source is not detected in the latest observation, taken with \chandra{} in 2022, and we infer an X-ray luminosity $\leq7\times10^{31}$\,\lum.
  The source spectral shape does not change over time and can be equally well fitted with each of the three models, with a best-fit photon index of 1.6 for the power law and best-fit temperatures of 10~keV for both the bremsstrahlung and the thermal plasma models. We confirm the previously detected period of 8649~s, ascribed to the binary orbital period, and found a cycle-to-cycle variability associated with this periodicity. No other significant pulsation is detected.}{Considering the source orbital period, luminosity, spectral characteristics, long-term evolution and strong cycle-to-cycle variability, we suggest that \src{} is a magnetic CV of the polar type.}
  \maketitle

\section{Introduction}
\label{sec:introduction}
    
Globular clusters (GCs) are the oldest stellar structures in the Milky Way, usually showing nearly spherical distributions, with tens of thousands of stars bound by gravity. Owing to their dense environments, GCs are good laboratories for stellar dynamics and host abundant populations of compact objects, often segregated in their core and formed dynamically \citep[e.g.][]{Pooley2003}. Examples include low-mass X-ray binaries (LMXBs), cataclysmic variables (CVs), recycled millisecond pulsars, and -- perhaps -- even intermediate mass black holes \citep{MeylanHeggie1997,Strader+2012,Lugaro+2013}. As the remnants of the least massive stars, which are the most abundant types of stars, white dwarfs (WDs) should be the most common compact objects in GCs. The identification of isolated WDs in GCs is a challenging endeavour due to their intrinsic faintness. On the other hand, the observation of accreting WDs in the UV and X-rays, their characteristics and population ratio with respect to other X-ray emitters are important diagnostics of the cluster evolution and tests of stellar dynamics.
    
47\,Tucanae (NGC\,104, 47\,Tuc for brevity) is one of the brightest and most massive GC in the Milky Way\footnote{A list of Milky Way GCs can be found at \url{https://physics.mcmaster.ca/~harris/mwgc.dat}}. It has a mass of $\simeq 7 \times 10^5$ M$_{\odot}$ \citep{Marks2010}, an age of 12-13 Gyr \citep{Zoccali+2001,GarciaBerro2014,Thompson+2020}, and various distance estimates: $4.521\pm0.031$~kpc \citep{BaumgardtVasiliev2021}, $4.45\pm0.01\pm0.12$~kpc \citep{Chen2018}, and $4.47\pm0.01\pm0.08$~kpc \citep{Simunovic2023}; for this work we shall simply assume 4.5~kpc. Several hundreds of CVs are expected to be present in 47~Tuc, as results of both the evolution of primordial binaries and dynamical encounters/three-body interactions \citep{PooleyHut2006,BelloniRivera2021}. Data taken with the \textit{Chandra X-ray Observatory} \citep{Weisskopf2000} of the core of 47\,Tuc showed the presence of more than one hundred faint X-ray sources within a few arcmin \citep{Edmonds2003b,Heinke2005}. About one-third were tentatively identified with CVs and many were confirmed with subsequent optical/ultraviolet observations \citep[see e.g.][]{Edmonds2003a,Bhattacharya2017,Rivera2017}. Recently, 47 Tuc has also been observed during the calibration phase of the \textit{extended Roentgen Survey with an Imaging Telescope Array} \citep[\textit{eROSITA},][]{Predehl2021}, where about 888 point-like sources were detected, with a handful identified as CVs \citep{Saeedi2022}.

\srcextended{} (W2 hereafter) was first identified as a possible CV candidate (or an enshrouded millisecond pulsar) by \citet{Grindlay2001}, who compared \chandra{} and \rst{} \citep{Truemper1982} data of 47 Tuc. \citet{Edmonds2003a,Edmonds2003b}  studied the optical counterparts of all X-ray sources of 47 Tuc with the \textit{Hubble Space Telescope} (\textit{HST}) and found for \src{} a V magnitude of 21.50 and U--V and V--I colours of 0.04 mag and 1.91 mag, respectively. The location of this source in the U,U--V colour-magnitude diagram was blueward of the main sequence, but close to the main sequence in the V,V--I colour-magnitude diagram. They concluded that the source might have had a relatively faint accretion disc, with the optical magnitude dominated by the companion star. They also suggested that the system had a low accretion rate and estimated an X-ray period of $\sim$6.3~hr, different from the optical periods of 2.2~hr, 5.9~hr, and 8.2 hr \citep{Edmonds2003b}.
The spectral analysis of the source was carried out by \citet{Heinke2005} on 2000 and 2002 \chandra{} data, resulting in X-ray luminosities in the energy range 0.5--6\,keV of $\sim8 \times 10^{31}$\,\lum{} and $\sim16 \times 10^{31}$\,\lum{}, for the two epochs, respectively.

\citet{Israel2016} unambiguously classified \src{} as a CV  based on the timing analysis of \chandra{} ObsIDs.~953, 955, and 2735--8, that revealed an eclipse with a period of 8649\,s (2.4\,hr), ascribed to the binary orbit. Their analysis did not confirm the X-ray period of $\sim$6.3~hr derived by \citet{Edmonds2003b}.
\citet{Rivera2017} further studied the \textit{HST} data of 47 Tuc in the near ultra-violet (NUV) and optical wavebands. Using the ultraviolet filter $U_{300}$, they did not find any periodicity of the source %(i.e. a filter for a wavelength of 300 nm, which is in the NUV range),
for any of the periods previously proposed in the literature. However, the authors selected only high-quality photometric data resulting in a sparse sampling over 8 hr of observation, which could well have missed the 8 min long eclipse. Notwithstanding, the light curve showed a variation of 3 mag during the time of the observation. On the other hand, the period of 2.4\,hr was confirmed in the more intensive coverage with the $R_{625}$ filter, with the appearance of a deep eclipse-like feature, hence confirming the optical counterpart to \src{}. The source was classified as a candidate magnetic CV, very likely an intermediate polar (IP), where the variations in magnitudes were attributed to either precession of the accretion disc, or changes of its thickness, or high/low state transitions. The most recent search for periodicity from the source was performed by \citet{Bao2023}, who found a first period of 8646.78~s, close to the one of \citet{Israel2016} and ascribed it to the orbital period, and a second one of 3846.15~s, interpreted as the spin period of the WD, supporting the IP scenario.  Unfortunately, due to the source density in the central region of 47 Tuc  (within a radius of 1.7$^\prime$), \src{} is not resolved by  \ero{} \citep{Saeedi2022}.

In this manuscript, we aim to resolve the controversy on the presence and the nature of the X-ray periods of \src{} and classify the source through simultaneous use of X-ray spectroscopic and timing analysis techniques. We use all available \chandra{} data, including two new observations taken in 2022 (see Tab.~\ref{tab:logChandra}). We also revisit \ero{} Early Data Release (EDR) and  detect the source for first time using this dataset, thanks to its period, even if it cannot be spatially resolved from nearby sources.  Data reduction, spectroscopic and timing analyses are described in Sect.~\ref{sec:data_reduction}, \ref{sec:spectroscopic_analysis} and \ref{sec:timing_analysis}, respectively. The nature of the source is discussed in Sect.\ref{sec:discussion} and our conclusions are presented in Sect.\ref{sec:conclusions}.

\section{Data reduction}
\label{sec:data_reduction}

\subsection{\chandra{}}
\label{sec:data_reduction_chandra}

\begin{figure*}
\resizebox{\hsize}{!}
{\includegraphics[]{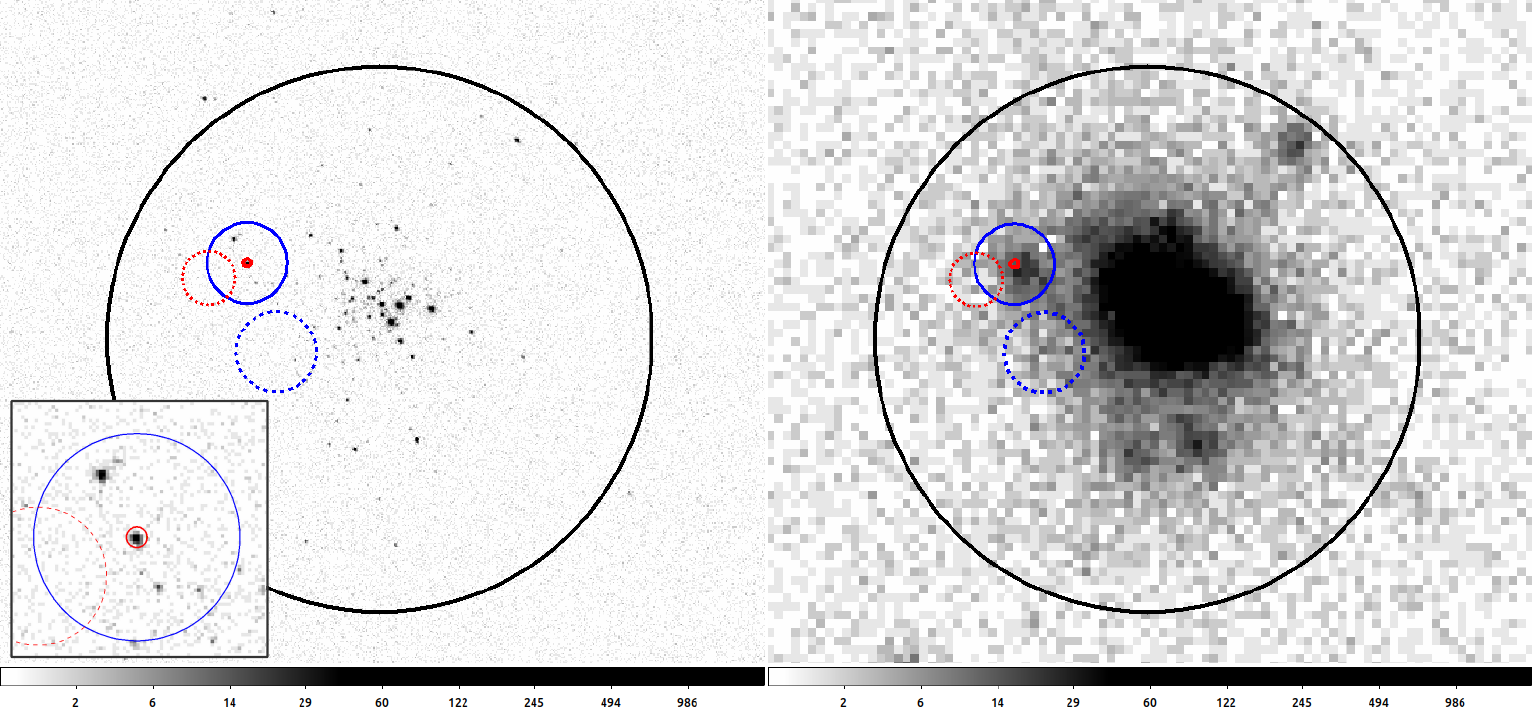}}
\caption{\chandra{} (left panel, ObsID.~2735) and \ero{} (right panel,1-2 Nov 2019 observations) images of the field of view of 47 Tuc. Solid and dashed circles represent the source and background extraction regions for \src{}, respectively, for \chandra{} (red) and \ero{} (blue). The black circle encloses the inner region of the cluster excluded in the \textit{eROSITA} analysis of \citet{Bao2023} and has a radius of 1.7$^\prime$. The box in the bottom left corner is a close-up of the \ero{} source region as seen by \chandra{}, to show the X-ray sources that are blended with \src{} in the \ero{} image.}
\label{fig:47_tuc_zoom_source}
\end{figure*}

All available observations of the Galactic GC 47 Tuc in the \chandra{} Data Archive span roughly 15 years, from early 2000 to 2015, with the addition of two recent observations, taken in January 2022. For this work we considered all the observations taken with ACIS-S or ACIS-I, as reported in Table~\ref{tab:logChandra}. A \chandra{} image of 47 Tuc is shown in Fig.~\ref{fig:47_tuc_zoom_source}, left panel.

All the data sets were reprocessed with the tool \texttt{chandra\_repro} in the \textit{Chandra Interactive Analysis of Observations software} (\textit{CIAO}, version 4.14), using calibration files CALDB 4.9.8. The target was identified by means of its sky coordinates, RA=$00^h24^m15.88^s$ and  DEC=$-72^{\circ}04^{\prime}36.38^{\prime\prime}$, in J2000 reference frame, as in \citet{Israel2016}.  Spectra and light curves were extracted using the \textit{CIAO} tool \texttt{specextract}. We used circular regions for both the source and the background, the former (1.8$^{\prime\prime}$) centred on the source position, the latter (10$^{\prime\prime}$) on a nearby region, which contained no other X-ray sources, on the same CCD. For each observation, we generated the source and background spectra, the redistribution matrix and the auxiliary response files (rmf and arf). Each spectrum was binned with a minimum number of 20 cts/bin. We used the same regions to extract the light curves, after having barycentred the observations at the source coordinates, with the JPL solar system ephemeris DE200. Amongst all the observations, two data sets (954 and 16528) were rejected, because the target was not on the illuminated CCDs. 

The source was not visible by eye in the latest observation ObsID.~26286. We ran the \textit{CIAO} tool \texttt{wavdetect} to see if the algorithm could detect it. We first created exposure-corrected images and exposure maps of the observation using \texttt{fluximage}, in the broad energy band (0.5--7 keV), where the source flux peaks \citep[e.g.][]{Heinke2005}, with the option \texttt{psfecf}=0.99. We then ran \texttt{wavdetect}, choosing the scales of 1 and 2 to enhance the detection of point-like sources. Seeing as no source was detected at the position of \src{}, we used the tool \texttt{aplimits} to obtain an upper limit on the count rate. For a false detection probability of 0.1 (corresponding to 90\% confidence level, c.l.), a probability of missed detection of 0.5, and a background rate of $7.5\times10^{-3}$ cts s$^{-1}$, estimated from the circular background region of all other observations, we obtained an upper limit of $1.2\times10^{-3}$ cts s$^{-1}$.

\subsection{eROSITA}
\label{sec:data_reduction_erosita}

\ero\ observed 47 Tuc during the calibration phase on November 1-2, 2019 (ObsIDs.~700011, 700163, 700013, 700014) and November 19, 2019 (ObsIDs.~700173-175), for a total of $\sim$101~ks and $\sim$25~ks, respectively. Another observation was carried out on September 28, 2019 (ObsID.~700012), but we excluded it since no information about the satellite orbit\footnote{\url{https://erosita.mpe.mpg.de/edr/eROSITAObservations/OrbitFiles}.} is available before October 2019 and no barycentric correction could be applied to the data. 

To extract \ero\ data from the EDR observations, we used the \ero\ Standard Analysis Software System \citep[\texttt{eSASS}, see][]{Brunner2022} available on the EDR webpage\footnote{\url{https://erosita.mpe.mpg.de/edr/DataAnalysis/esassinstall.html}.}. Following the online guide to \texttt{eSASS}\footnote{\url{https://erosita.mpe.mpg.de/edr/DataAnalysis/esasscookbook.html}.}, we selected all the events in the 0.2--5~keV band with valid patterns (PATTERN=15) within the nominal field of view (FLAG=0xc0008000). In Fig.~\ref{fig:47_tuc_zoom_source} (right panel) we show a close-up view of the eROSITA field during the November 1-2 observations. 
The solid blue circle denotes the 15$^{\prime\prime}$-radius region we used to extract the source events. Note that \ero{} is not able to resolve \src{} from the other X-ray sources within the extraction area, which are instead resolved by \chandra{} (see box in the bottom left corner of Fig.\ref{fig:47_tuc_zoom_source}). In particular, another CV \citep[W1, see e.g.][]{Heinke2005}, with luminosity comparable to \src{}, falls within the \ero{} extraction region. This second CV was detected in the 2000 and 2002 \chandra{} data sets by \citet{Heinke2005}, while a visual inspection of the 2014 and 2022 data sets revealed it to be switched off. We have no means to say whether W1 is also present in the \ero{} observations.

\section{Spectroscopic analysis}
\label{sec:spectroscopic_analysis}

Spectral analysis was performed with the X-Ray Spectral Fitting Package \textit{Xspec} \citep{Arnaud1996}, version 12.9.1, using abundances from \citet{Wilms2000} and cross-sections from \citet{Verner1996}. At first, we considered only the seven observations (ObsIDs.\,955, 2735, 2736, 2737, 2738, 15747, 16527), where the source spectrum has at least five data points in the range 0.5--6\,keV, suitable for a fit with $\chi^2$ statistics. Above 6\,keV the spectrum is dominated by the background. Following the  analysis by \citet{Heinke2005}, we fit them separately with three different emission models: a power law (PL), a thermal Bremsstrahulung (TB), and a \texttt{vmekal} \citep{Mewe1991} model. Statistical errors are at 90\% c.l., unless specified otherwise.

We considered two different components for the interstellar absorption. The first component ($N_\mathrm{H}^\mathrm{GAL}$, modelled with \texttt{TBabs}) accounts for the total Galactic absorption in the direction of 47 Tuc and was fixed to $5.5 \times 10^{20}$ cm$^{-2}$, according to the estimate provided by the HEASARC $N_\mathrm{H}$ calculator tool\footnote{https://heasarc.gsfc.nasa.gov/cgi-bin/Tools/w3nh/w3nh.pl}, assuming Solar abundances. The second component ($N_\mathrm{H}^\mathrm{47 Tuc + local}$, described by the \texttt{TBvarabs} function) represents both the interstellar absorption within the cluster and  local to the source. In this case, we fixed the chemical abundances to those derived from \citet{Thygesen2014} for 47 Tuc, namely 33\% of solar abundances for C, N, O, and Ne, 27\% for Na and Al, 46\% for Mg, 35\% for Si, S, and Ar, 34\% for Ca, 17\% for Fe, and 13\% for Ni. 

For all three models, the local absorption $N_\mathrm{H}^\mathrm{47 Tuc + local}$ was consistent in all the spectra, with the exception of ObsIDs.~15747 and 16527, where it was unconstrained. Hence, we tied the $N_\mathrm{H}^\mathrm{47 Tuc + local}$ values and fitted the seven observations again. The fit converged to $N_\mathrm{H}^\mathrm{47 Tuc + local}\sim2 \times 10^{21}$ cm$^{-2}$. Detailed tables with the best-fit results for both fits and for each model are reported in Appendix \ref{sec:appendixB}.

The best-fit power law photon index and the temperatures of the TB and \texttt{vmekal} models were consistent for all observations, so that we also tied them together to get more stringent constraints. We obtained a best-fit photon index $\Gamma = 1.53^{+0.13}_{-0.12}$ and best-fit temperatures of $11^{+7}_{-4}$\,keV for both thermal models. According to the fit statistics, all models provide an equally good description of the spectra.

As a side note, for the \texttt{vmekal} model, we considered the same abundances we used for the local absorption component \citep{Thygesen2014}. The total spectrum was then computed by interpolating on a pre-calculated table (parameter \texttt{switch$=$1}). Setting the chemical abundances to those of \citet{Heinke2005}, leads to little or no difference in the best-fit values. We also note that there is little to no difference in the fit residuals between the two thermal models (Fig.\ref{fig:FitTotPwrlaw}), confirming that the two models are equivalent with respect to the quality of the present data sets. We do not include \ero{} data in the spectral analysis, because it is not possible to extract a clean source spectrum, not contaminated by the nearby bright X-ray source W1 (cf. Sect.\,\ref{sec:data_reduction_erosita} and Fig.\,\ref{fig:47_tuc_zoom_source}). This will not be a problem for the timing analysis (Sect.\,\ref{sec:timing_analysis}).

\subsection{Combined data}

\begin{figure}
 \resizebox{\hsize}{!}
{\includegraphics[]{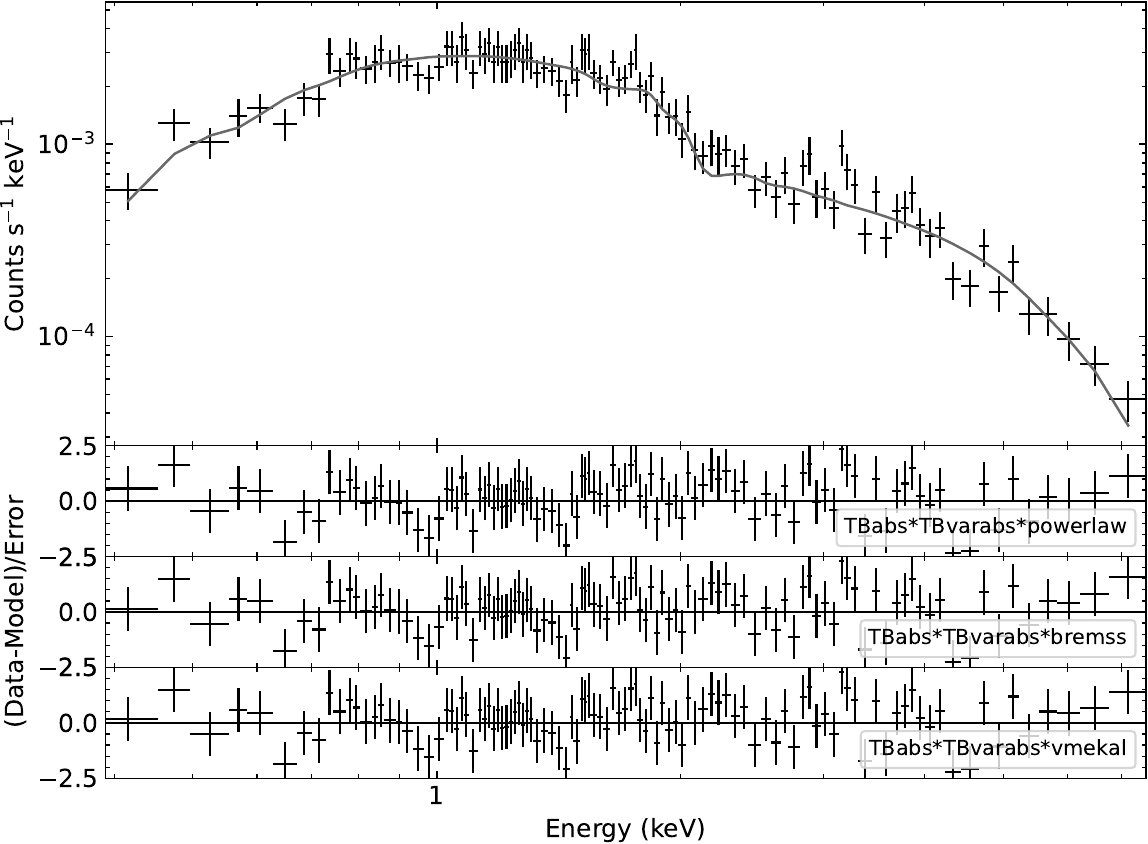}}
    \caption{Summed spectra from all the \chandra\ data sets in which the source was detected, together with the best-fit power-law model (top panel), and residuals in units of $\chi^2_{red}$ in the case of the best-fit power-law model (second panel), Bremsstrahlung model (third panel), and \texttt{vmekal} model (bottom panel).}
    \label{fig:FitTotPwrlaw}
\end{figure}

The consistency of the best-fit parameters of the previous fits demonstrates that \src{} does not change its spectral state over time, in spite of changes in flux (the normalisation is changing). Therefore, we decided to combine the spectra of all observations in which the source was detected (Table \ref{tab:logChandra}), to further constrain the best-fit parameters. To combine the spectra, we used the \textit{CIAO} tool \texttt{combine\_spectra}, which sums the counts of each corresponding spectral bin of different spectra and returns a total spectrum. In this way, we increased the number of counts in each spectral bin and achieved a better signal-to-noise ratio (S/N). %This operation could also allow to better discern whether one model is better than the other. 
The combined spectrum was again grouped with 20 cts/bin and fitted  with the same models as above. Results of the fits are shown in Table \ref{tab:FitTot} and in Fig. \ref{fig:FitTotPwrlaw}, with fluxes and luminosities in the 0.3--10 keV energy range.
%\ref{tab:FitSum8}.
The fits returned a $\chi^2_{red}$(d.o.f) of 0.94(97) for the PL and of 0.90(97) for both the TB and the \texttt{vmekal} models, with null hypothesis probability (n.h.p.) values of 0.64, 0.75, and 0.74, respectively. The best-fit values of the parameters $N_\mathrm{H}^\mathrm{47 Tuc + local}$, $\Gamma$ and $kT$ were all consistent with those obtained in the previous fits. In all cases, the unabsorbed flux is (6--7)$\times 10^{-14}$ \flux{}, which results in X-ray luminosities in the range $L_{\rm X} = (1.5-1.7) \times 10^{32}$ \lum{}, for a distance $d=4.5$~kpc.

\begin{table*}
\centering			
\caption{Best-fit values of the spectrum obtained by combining all the data sets.} 
\begin{tabular}{lccc}
\hline\hline\noalign{\smallskip}
Model & PL & TB & \texttt{vmekal}\\  \noalign{\smallskip}\hline\noalign{\smallskip}
$N_\mathrm{H}^\mathrm{47 Tuc + local}$ ($10^{21}$\,cm$^{-2})$ & 1.8$\pm$0.8 & 1.0$\pm$0.6 & 1.0$\pm$0.6 \\ \noalign{\smallskip}
$\Gamma$ & 1.55$^{+0.10}_{-0.09}$ & -- & -- \\   \noalign{\smallskip}
$kT$ (keV) & -- & $10_{-2}^{+5}$ & $10_{-2}^{+4}$\\ \noalign{\smallskip}
Norm (10$^{-6}$)\tablefootmark{a} & $8.5 \pm 0.8$ & $11.1 \pm 0.5$ & $34_{-1}^{+2}$\\  \noalign{\smallskip}
$F_\mathrm{abs}$ ($10^{-14}$\,\flux{}) &	5.8$\pm$0.3 & $5.4_{-0.5}^{+0.3}$ & $5.5_{-0.7}^{+0.3}$\\   \noalign{\smallskip}
$F_\mathrm{unabs}$ ($10^{-14}$\,\flux{})  &	$6.7\pm0.3$ & $6.0^{+0.4}_{-0.3}$ & $6.1^{+0.3}_{-0.4}$\\   \noalign{\smallskip}
$L_\mathrm{X}$ ($10^{32}$\,\lum{})
& $1.62\pm0.07$ & $1.45^{+0.10}_{-0.07}$ & $1.48^{+0.07}_{-0.10}$\\  \noalign{\smallskip}
$\chi^2_{red}$(d.o.f.) & 0.94(97) & 0.90(97) & 0.90(97)\\   \noalign{\smallskip}
n.h.p. & 0.64 & 0.75 & 0.74\\   \noalign{\smallskip}\hline\hline
\end{tabular}
\tablefoot{
\tablefoottext{a}{In units of photons keV$^{-1}$ cm$^{-2}$ s$^{-1}$ at 1 keV for the PL model and of cm$^{-5}$ for the TB and the \texttt{vmekal} models.
}}
\label{tab:FitTot}
\end{table*}

\begin{figure}
 \resizebox{\hsize}{!}
    {\includegraphics[]{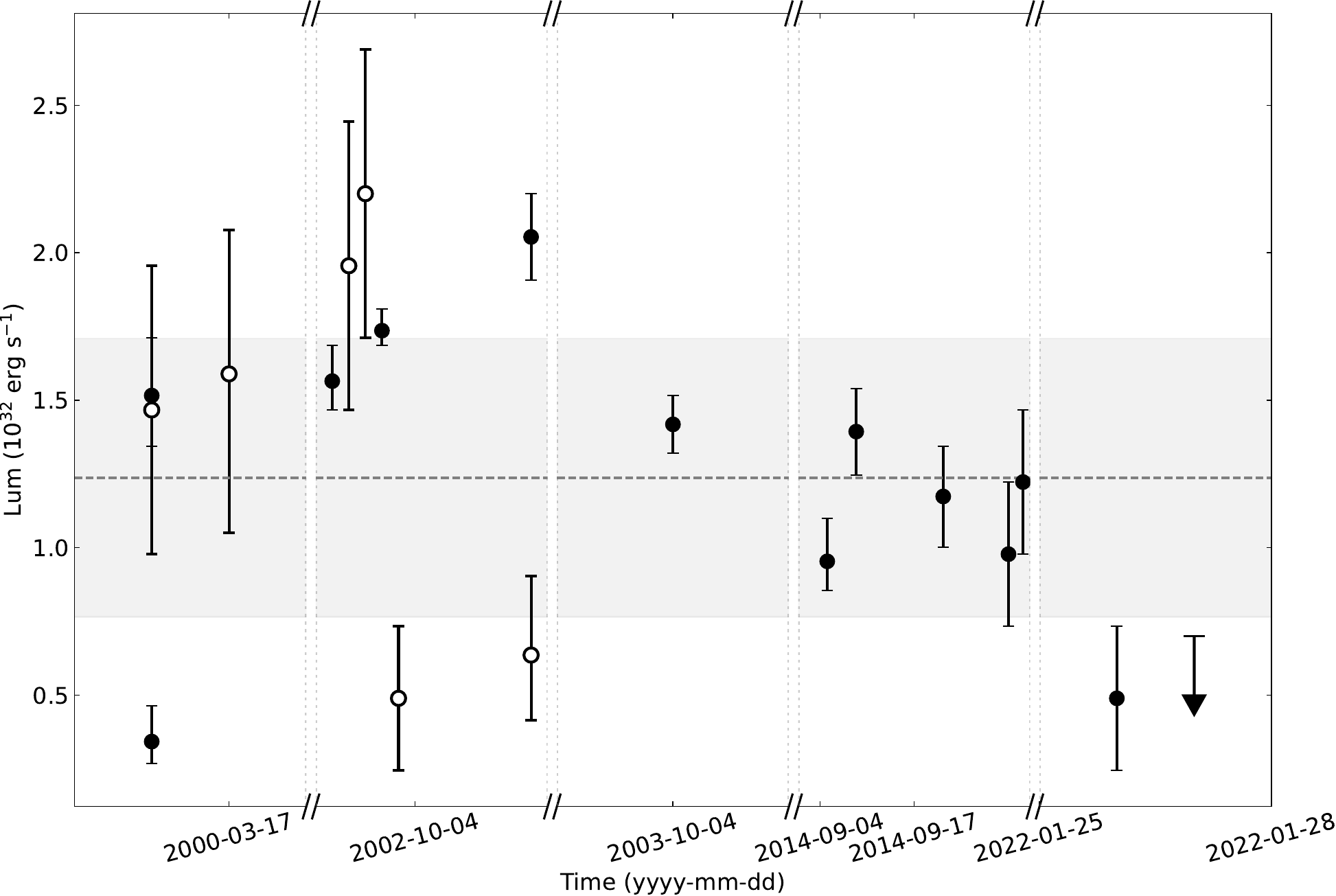}} 
    \caption{Light curve of \src{}. Empty data points are for those observations with exposures shorter than the orbital period; the black arrow represents the upper limit at 90\% c.l. on the luminosity of the latest \chandra{} observation (ObsID.\,26286); the grey dotted line and area represents the average luminosity of the source and its uncertainty at 90 \% c.l., respectively.  
    }
    \label{fig:lc_total}
\end{figure}

Once the best-fit parameters were constrained, we explored the long-term variability of the source, studying the changes in flux over time. To do so, we used the TB model and derived the fluxes for those observations excluded from the previous analysis. We imposed $N_\mathrm{H}^\mathrm{47 Tuc + local}=1.1 \times 10^{21}$\,cm$^{-2}$ and $kT=10$\,keV (see Table \ref{tab:FitTot}) and fitted the data, leaving only the normalisation free to vary. For the latest \chandra{} observation, for which we do not have a spectrum, we derived upper limits on the flux and luminosity from the upper limit on the count rate estimated in Sect.\,\ref{sec:data_reduction_chandra}. We used the online tool WebPIMMS\footnote{\url{https://heasarc.gsfc.nasa.gov/cgi-bin/Tools/w3pimms/w3pimms.pl}.} with the best-fit parameters of the TB model and obtained an X-ray luminosity $\leq7\times10^{31}$ \lum{}, in the energy range 0.3--10\,keV, at 90\% c.l. All resulting fluxes and luminosities are shown in Table~\ref{tab:lum_brems} and in Fig. \ref{fig:lc_total}. 
In most observations, the source flux is in the range $F_\mathrm{abs} \sim (4-9) \times 10^{-14}$ \flux{}, corresponding to a luminosity $L_\mathrm{X}\sim (1-2)\times10^{32}$ \lum{}, while in a few it is significantly lower. Some observations (marked with an asterisk in Table \ref{tab:lum_brems}) are shorter than the orbital period and their luminosities might be unreliable. 

\begin{table}
\centering
\caption{\src{} unabsorbed fluxes and luminosities (0.3--10 keV) for each observation\tablefootmark{a}. An asterisk marks those observations with exposure times shorter than the orbital period.} 
\begin{tabular}{cccc} 
\hline\hline\noalign{\smallskip}
Obs. ID &   $F_\mathrm{unabs}$     & $L_\mathrm{X}$           \\
        &   (10$^{-14}$ \flux{})    & ($10^{32}$ \lum)  \\ 
\hline\noalign{\smallskip}
78*      & 6$\pm$2               & 1.5$\pm$0.5           \\ \noalign{\smallskip}
953     & 1.4$^{+0.5}_{-0.3}$    & 0.34$^{+0.12}_{-0.07}$ \\ \noalign{\smallskip}
955     & 6.2$^{+0.8}_{-0.7}$    & 1.5$\pm$0.2           \\ \noalign{\smallskip}
956*     & 6.5$^{+2.0}_{-2.2}$    & 1.6$\pm$0.5           \\ \noalign{\smallskip}
2735    & 6.4$^{+0.5}_{-0.4}$    & 1.5$^{+0.12}_{-0.09}$ \\ \noalign{\smallskip}
3384*    & 8$\pm$2               & 2.0$\pm$0.5           \\ \noalign{\smallskip}
2736    & 5.8$\pm$0.4           & 1.42$\pm$0.09         \\ \noalign{\smallskip}
3385*    & 9$\pm$2               & 2.2$\pm0.5$    \\ \noalign{\smallskip}
2737    & 7.1$^{+0.3}_{-0.2}$    & 1.73$^{+0.07}_{-0.05}$ \\ \noalign{\smallskip}
3386*    & 2$\pm$1               & 0.5$\pm$0.2           \\ \noalign{\smallskip}
2738    & 8.4$\pm$0.6           & 2.0$\pm$0.1           \\ \noalign{\smallskip}
3387*    & 2.6$^{+1.1}_{-0.9}$    & 0.6$^{+0.3}_{-0.2}$    \\ \noalign{\smallskip}
16527   & 3.9$^{+0.6}_{-0.4}$    & 0.95$^{+0.14}_{-0.09}$ \\ \noalign{\smallskip}
15747   & 5.7$\pm$0.6           & 1.4$\pm$0.1           \\ \noalign{\smallskip}
16529   & 4.8$\pm$0.7           & 1.2$\pm$0.2           \\ \noalign{\smallskip}
17420   & 4$\pm$1               & 1.0$\pm$0.2           \\ \noalign{\smallskip}
15748   & 5$\pm$1               & 1.2$\pm$0.2           \\ \noalign{\smallskip}
26229   & 2$\pm$1               & 0.5$\pm0.2$    \\  
\noalign{\smallskip}
26286 & $\leq$2.8\tablefootmark{b} & $\leq$0.7\tablefootmark{b} \\
\noalign{\smallskip}
\hline\hline
\end{tabular}
\tablefoot{
\tablefoottext{a}{Fluxes and luminosities have been obtained for the thermal Bremsstrahlung model, having $N_\mathrm{H}^\mathrm{47 Tuc + local}$ and $kT$ fixed at $1.1 \times 10^{21}$ cm$^{-2}$ and 10 keV, respectively.
}
\tablefoottext{b}{Upper limit at 90\% c.l.}
}
\label{tab:lum_brems}
\end{table}

\section{Timing analysis}
\label{sec:timing_analysis}

\begin{figure*}
\centering
\begin{tabular}{c@{\hspace{1pc}}c@{\hspace{1pc}}c@{\hspace{1pc}}c}
    {\includegraphics[height=6.75truecm]{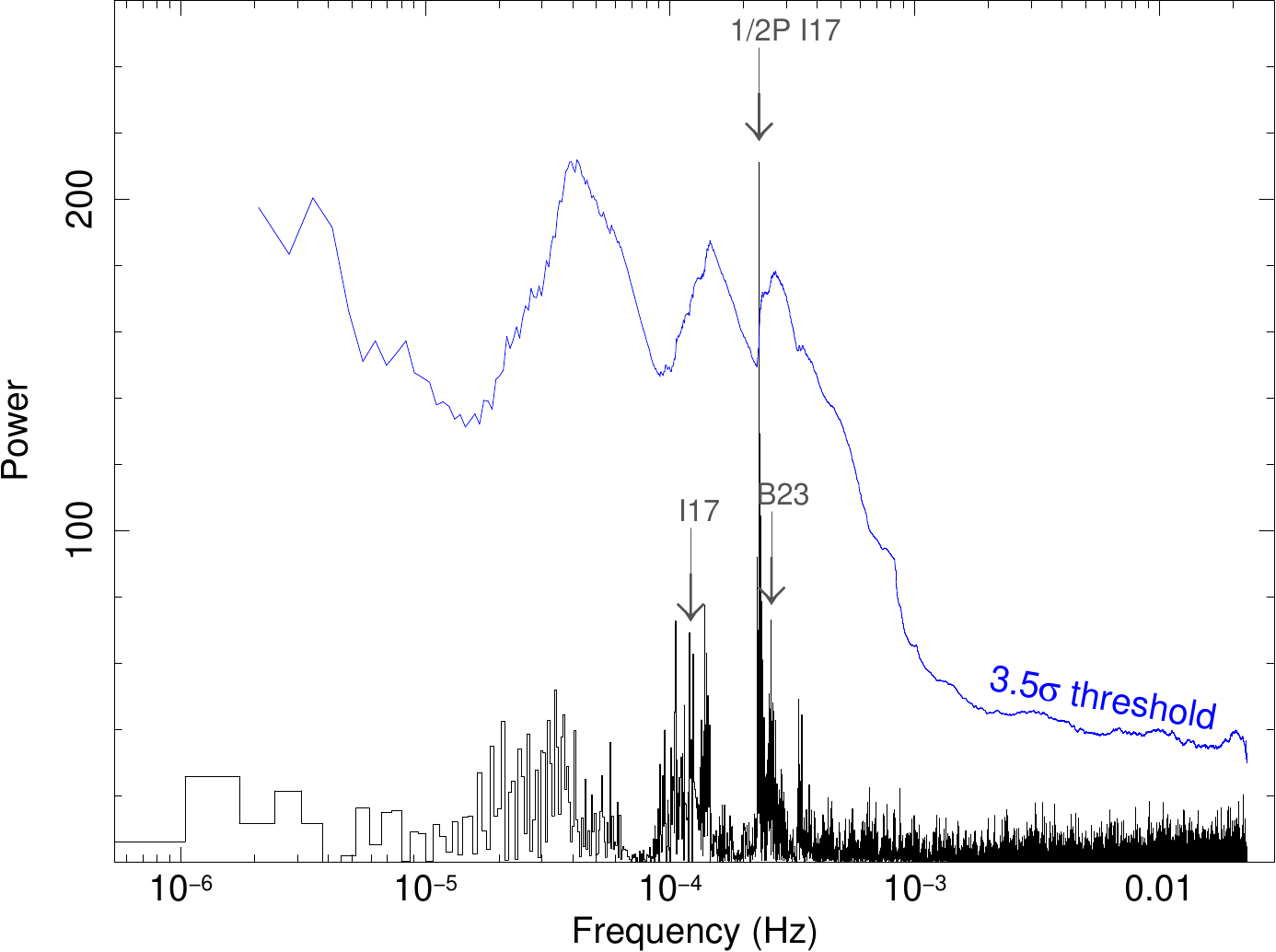}} \quad
\includegraphics[scale=0.35]{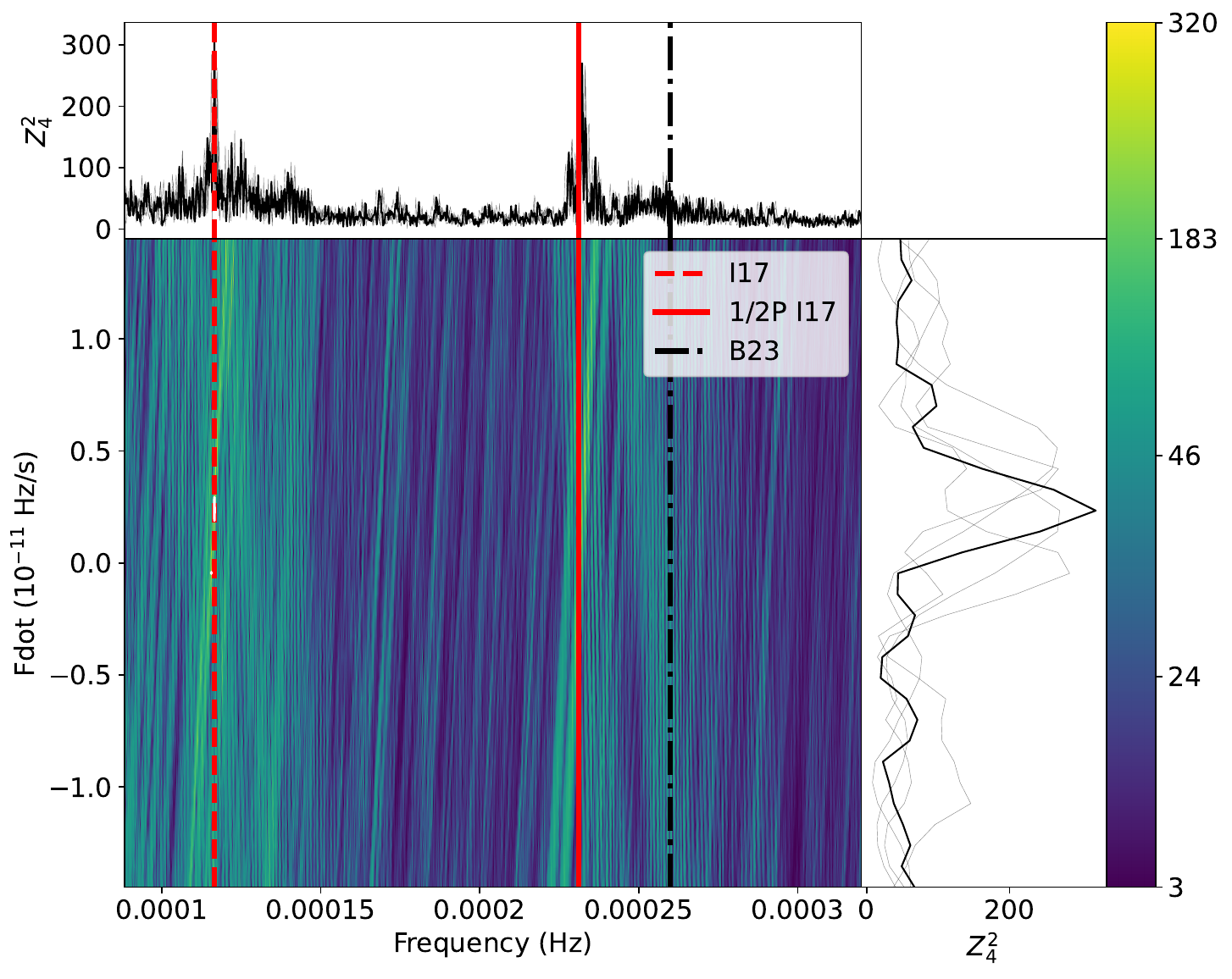} 
\end{tabular}
    \caption{Left: \chandra\ 0.2-10~keV power density spectrum (PDS) of \src\ by using the 2002 datasets 2735--8 and one single interval. The 3.5$\sigma$ detection threshold is also shown and marked by the blue solid line. The PDS is complex and only one peak is above the threshold, corresponding to the harmonic of the 8649s period (both the fundamental and the harmonic are marked by I17 and 1/2P I17, respectively). The period suggested by \citet{Bao2023} is indicated as B23. Right: $Z^2_N$ search with $N=4$ harmonics of \chandra{} ObsIDs.\,2735--2738. The fundamental and the first harmonic of the period at $P\simeq8650$~s are marked with red dashed and solid lines, respectively, while the period proposed by \citet{Bao2023} is marked with a black dash-dotted line.} 
    \label{dps}
\end{figure*}

\citet{Israel2016} found a phase-coherent solution for \src{} in ObsIDs.~2735--8 corresponding to a period of 8649$\pm$1~s. The eclipse profile was asymmetric and showed a total eclipse lasting about 8~min. The modulation was also present in ObsIDs.~953 and 955, though the poor statistics did not allow an independent period value to be inferred. Among the new available archival observations we focused on ObsID.~15747--8, 16527 and 16529 where the source is detected and the relatively long exposures ensure good statistics. A phase-fitting procedure was applied revealing a period of 8653$\pm$3~s, in agreement with that in the CATS@BAR catalogue although with a larger uncertainty. Consequently, we keep the period reported in the catalogue as the reference one in the subsequent analysis.

\begin{figure}
\resizebox{\hsize}{!}
{\includegraphics[]{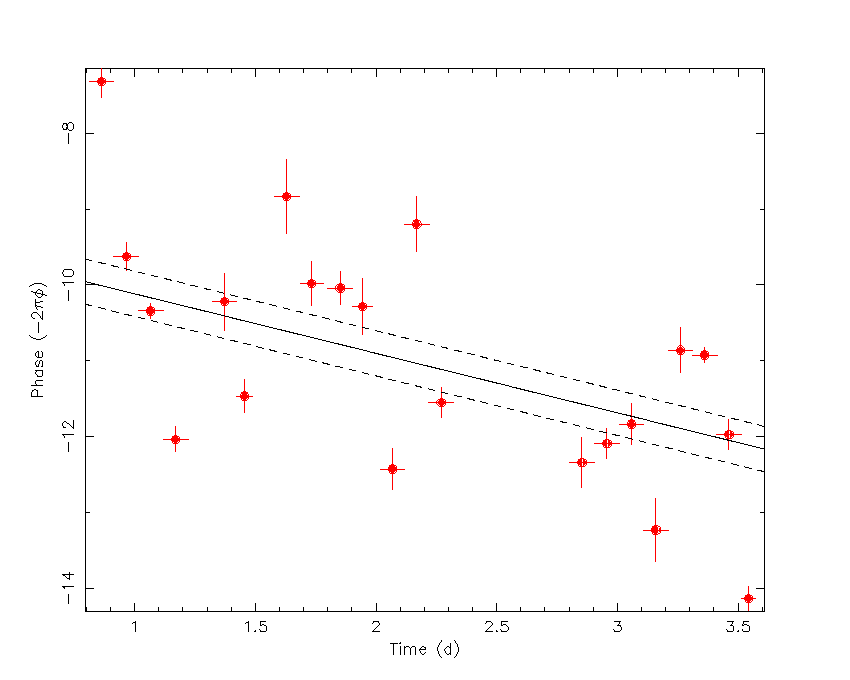}}
\caption{Phase-fitting of the 3846.15\,s period of \citet{Bao2023}, where each data point corresponds to a duration of one orbital cycle (i.e. 8649\,s). The phases are highly scattered up to almost 40\% of the period, suggesting that this period is spurious and/or not coherent. The black solid and stepped lines mark the best fit linear component and its 1$\sigma$ uncertainty, respectively.}
\label{checkPspin}
\end{figure}

\begin{figure}
\centering
\resizebox{\hsize}{!}    {\includegraphics[]{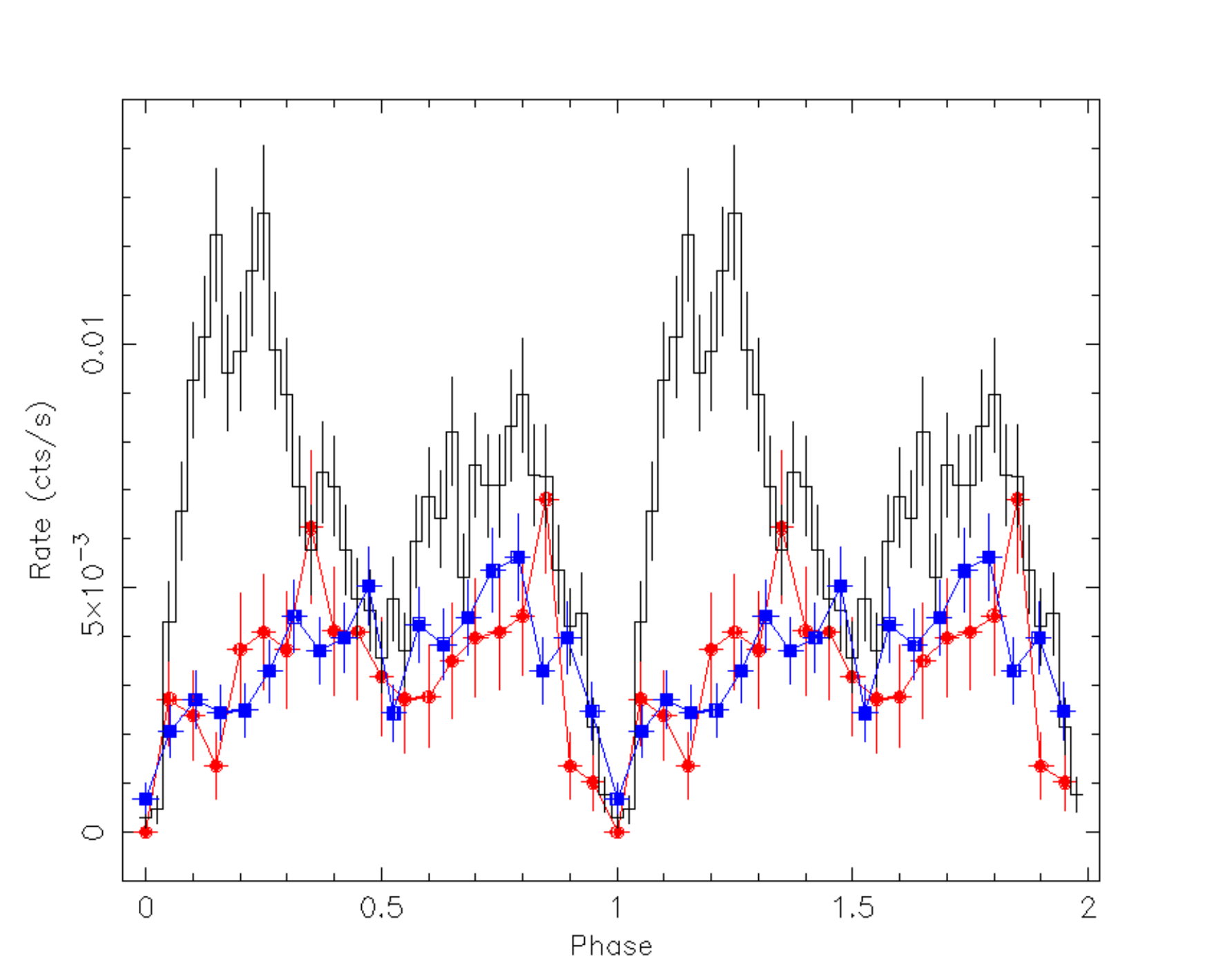}}\vspace{-5mm}
{\includegraphics[width=0.48\textwidth]{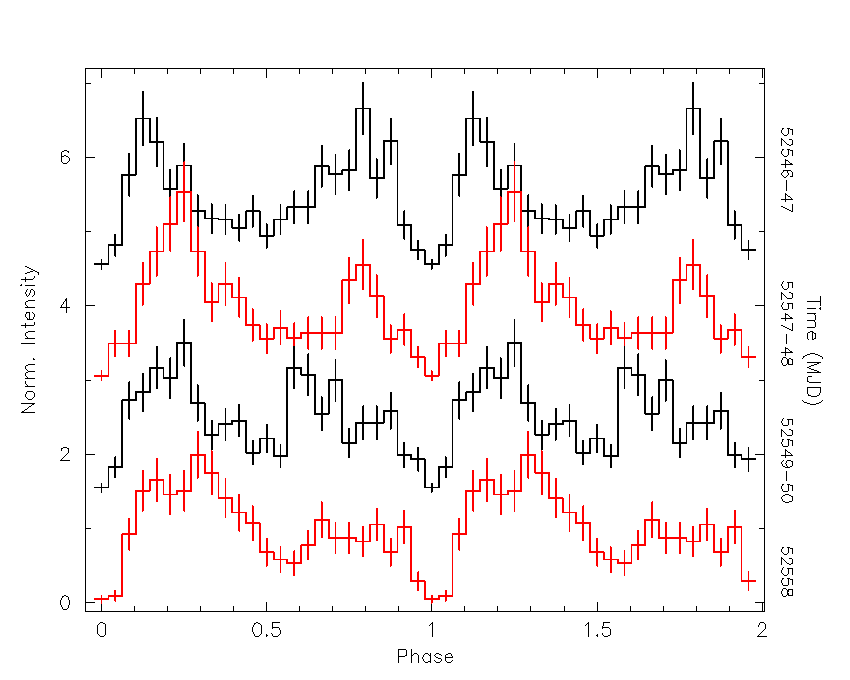}}
\caption{Top panel: \chandra\ ACIS eclipse profiles of \src{}, folded to the best period of $P=8649$~s, for the three different epochs where the modulation was detected: 2000 (red line and filled circles), 2002 (black lines), and 2014 (blue line and filled squares). A correction of about 20\% was applied to the 2000 datasets in order to take into account the difference in CCD efficiency between ACIS-I and ACIS-S. We used the reference epoch MJD 52551.999(1) (1$\sigma$ c.l.) for the 2002 datasets, while the 2000 and 2014 light curves have been arbitrarily shifted along the x-axis in order to align the minima, corresponding to the 8-min eclipse. All light curves have 20 phase-bins of $\sim$432\,s duration each. Bottom panel: Folded light curves of the four longest observations, carried out in 2002 a few days apart from each other. From top to bottom: ObsID.~2735, 2736, 2737 and 2738. The profiles are complex and variable on time scales of less than a day. 
}
    \label{profiles}
\end{figure}

\begin{figure}
\centering
\resizebox{\hsize}{!}    
{{\includegraphics[]{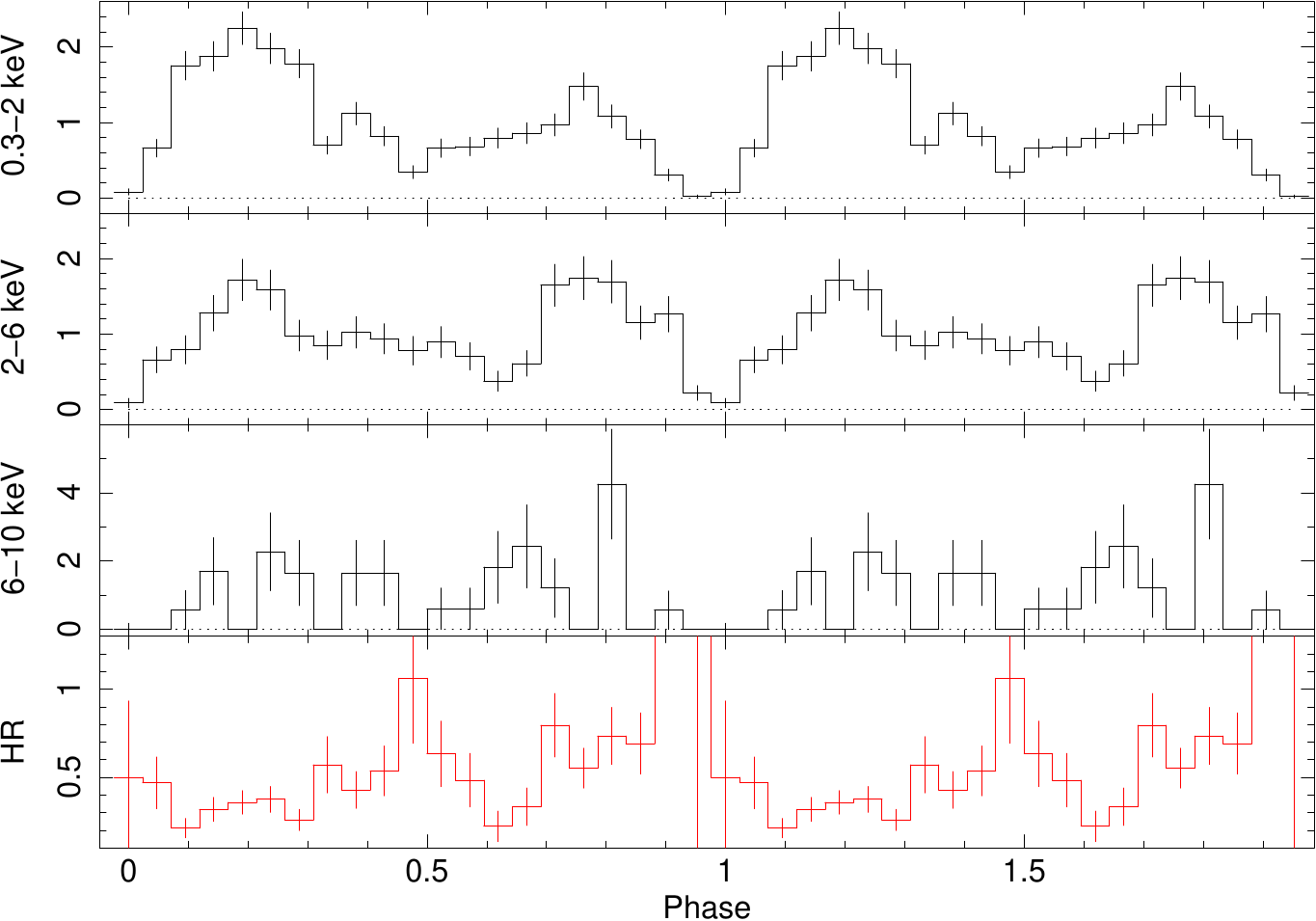}}}       
\caption{Energy-resolved profiles of the summed 2002 data sets, in units of normalised intensity, for the following energy ranges: 0.3--2~keV (first panel), 2--6~keV (second panel), 6--10~keV (third panel). Bottom panel: hardness ratio (2--6~keV)/(0.3--2~keV).   
    }
    \label{profiles_energyranges}
\end{figure}

\begin{figure}
\centering
\resizebox{\hsize}{!}
{\includegraphics[]{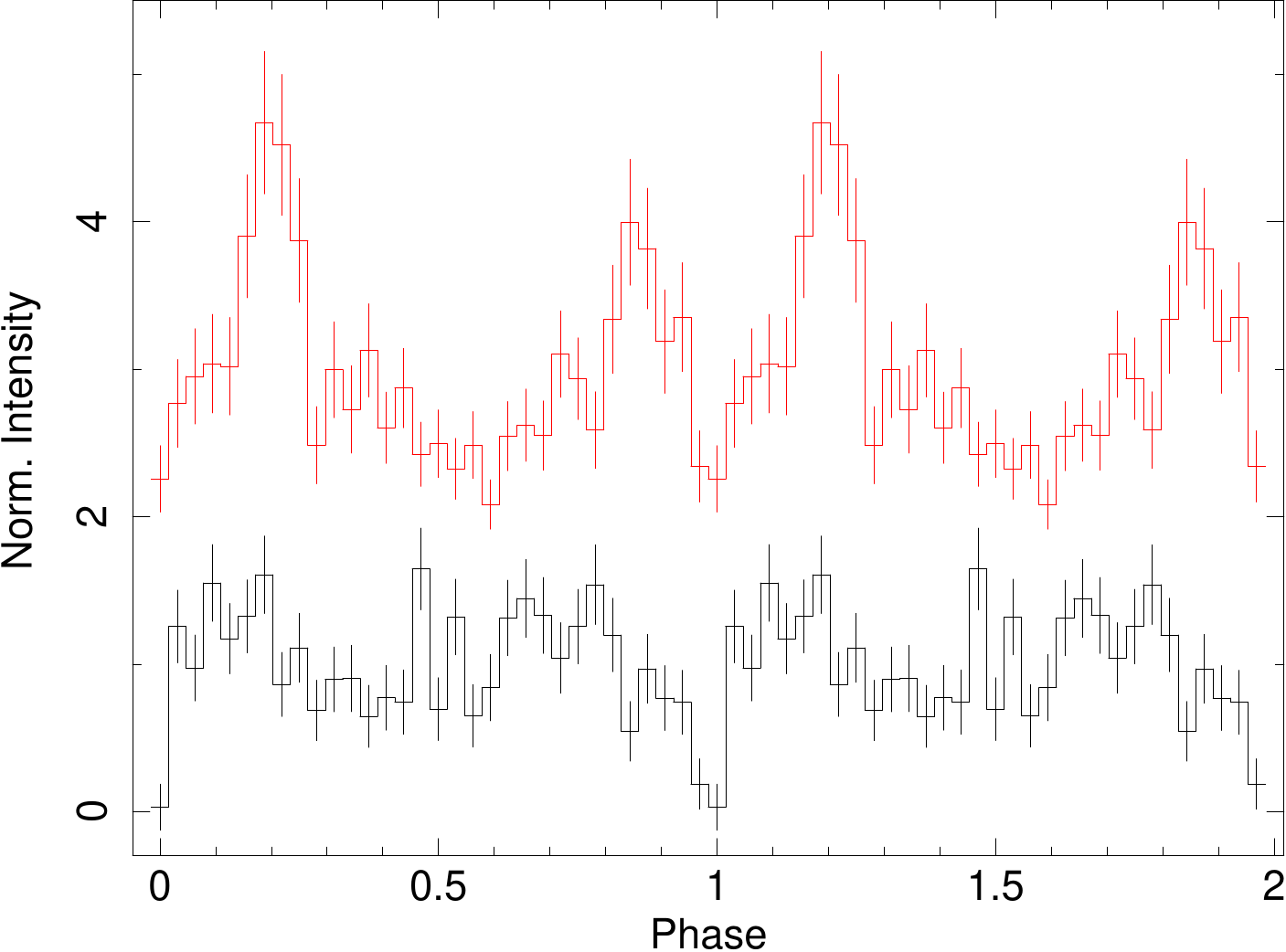}}
    \caption{Phase-folded profile ($P=8649$~s) in \ero{} data from the observations on November 1-2, 2019 (black profile, reference epoch MJD 58787.991(1), 1$\sigma$ c.l.) %MJDREF = 18787.98, P = 8649 s, newbins/interval = indef, bkg subtracted = 1.33e-2. 
     and on November 19, 2019 (red profile, shifted on the y-axis for better visualization). The latter has also been arbitrarily shifted along the x-axis to align the eclipse phase.%MJDREF = same as before, P = 8649 s, newbins/interval = indef, bkg subtracted = 1.33e-2.
    }\label{fig:eROSITAsignal}
\end{figure}

Recently, \citet{Bao2023} reported a second coherent signal, detected by a Gregory-Loredo algorithm \citep{Gregory1992}, corresponding to a period $P=3846.15$~s. Our PDS analysis of \chandra\ observations 2735--8, performed following \citet{Israel1996}, shows no significant peak above the 3.5$\sigma$ threshold at the corresponding frequency  (see Fig.\ref{dps}, left panel), although the high PDS peak at the frequency of the second harmonic of the 2.4\,hr modulation, at about 4325\,s, hampers the detection sensitivity around the peak itself. To further check for the presence of the reported modulation at $\sim3846$\,s, we also performed a Z$^2_n$ search \citep{Buccheri1983} for periodicity in the 0.1--0.6~mHz range, using the \texttt{henzsearch} tool included in the \texttt{HENDRICS} package \citep{Bachetti2018}. The right panel of Fig.~\ref{dps} %\ref{fig:Zsearch} 
shows the results of our analysis. Both the fundamental ($\nu\simeq116\,\mathrm{\mu Hz}$, $n=4$) and the second harmonic ($\nu\simeq231\,\mathrm{\mu Hz}$, $n=1$) are clearly detected, while the $P=3846.15$~s is below the 3.5$\sigma$ detection threshold. 

In order to further check the possibility that the peak at $P=3846.15$~s  might be a spin period we carried out a phase-fitting analysis by using time intervals of 2.4\,hr length (i.e. over one orbital cycle). If the  $P=3846.15$~s is real, we expect to see either a constant or a linear trend in the phases. In Fig.\,\ref{checkPspin} we show the results of this analysis ($\chi^2_\mathrm{red}$ of 40 for a linear component model) where it is evident that the phases are highly scattered up to almost 40\% of the period, suggesting that the peak is spurious and/or marks a variability with a low coherence level such as a QPO-like component. Regardless of the exact origin of the 3846.15~s peak, we can reasonably exclude that it is a strictly coherent signal and, therefore, that it is the spin period of the accreting WD. 

Fig.~\ref{profiles} shows the eclipse profile evolution as a function of time, for $P=8649$~s, for the three different epochs (2000, 2002, and 2014, left panel), while the bottom panel shows the eclipse profiles for the 2002 ObsIDs.2735--8. Note that the sparseness of the observations does not allow for a phase alignment of the three epochs. Hence, we shifted the profiles arbitrarily in order to have the centre of the eclipse aligned at the same phase. In Fig.~\ref{profiles_energyranges} we show the eclipse profiles of the summed 2002 spectra at different energy ranges (0.3--2~keV, 2--6~keV, and 6--10~keV) and the hardness ratio (2--6 keV/0.3--2 keV). The profiles clearly display a double hump, more evident at soft than hard energies, with the relative intensity of the humps changing both from epoch to epoch and within the same epoch, often on time scales less than a day. The hardness ratio is constant within the uncertainties, except for the eclipse phase, where the counts in all bands are consistent with zero.

To complete our analysis, we searched for the same periodicity of $P=8649$~s in the \ero{} data. Both the PDS and $Z^2_N$ search did not reveal any signal with a significance $\geq3.5\sigma$, but epoch folding at the period found in \chandra{} data returned the same eclipse profile at $P=8649$~s. Fig.~\ref{fig:eROSITAsignal} shows the phase-folded profiles of both \ero{} epochs, where the background has been inferred by also correcting for the expected count rate of the other bright source W1 within the source extraction region (see box in Fig.\ref{fig:47_tuc_zoom_source}) in the \ero\ energy band, based on the analysis of \citet{Heinke2005} of \chandra\ archival observations where W1 is detected. 
\src{}'s eclipse profile shape is consistent with the one found in the \chandra\ data (cf. Fig.~\ref{profiles}), ultimately proving the detection of the source. We also tried to fold \ero{} light curves at the period of $P\simeq3846.15$~s and obtained a profile similar to the one reported by \citet{Bao2023}. However, we also found that similar profiles are obtained with any period within the 3746--3946\,s range, and indeed at these frequencies both the Z$^2_n$ and PDS analysis shows an excess of power with respect to a pure white noise level (see Fig.\,\ref{dps}, right panel).

\section{Discussion}
\label{sec:discussion}

This work attempts at ascertaining the nature of the CV \src{} in the Galactic GC 47 Tuc. To do so, we carried out a comprehensive spectral and temporal analysis, using all available data sets of the \chandra{} Data Archive and six \ero{} EDR observations. 

We found that the X-ray spectrum of the source can be described equally well using three different models: a power law, a thermal bremsstrahlung, and an optically thin thermal plasma (\texttt{vmekal}). All fits need to include the absorption of the interstellar medium, in excess of $\sim1\times 10^{21}$ cm$^{-2}$. The power law model returned a best-fit photon index of $\Gamma=1.55$, while the thermal models resulted in the best-fit temperatures of $kT=10$ keV. The latter result is in agreement with those of \citet{Heinke2005} for the 2002 data set (ObsIDs.~2735--8, 3384--7), but not for the 2000 one (ObsIDs.~78, 953--6), which had a higher temperature of the \texttt{vmekal} model. However, it has to be noted that we only used one  (ObsID.~955) out of the five 2000 observations to determine the best-fit parameters, which resulted consistent with those of the 2002 data sets. 

The inferred unabsorbed fluxes are almost all consistent with each other within uncertainties, resulting in luminosities around the average value of $1.3\times10^{32}$ \lum{} (Fig.\,\ref{fig:lc_total}), typical of magnetic CVs \citep[e.g.][]{Mukai2017}. In the latest observations (ObsIDs.26229 and 26286), both taken in 2022, the source appears to have weakened, with a luminosity of $5\times10^{31}$ \lum{} for ObsID.26229 and an upper limit of $7\times10^{31}$ \lum{} for ObsID.26286.  

The period $P=8649$ s found by \citet{Israel2016} for the epochs 2000 and 2002 is confirmed also for the observations taken in 2014--15. The presence of a $\sim$8\,min eclipse strongly indicates that this period can be ascribed to the orbital period of the system, placing \src{} among the CVs in the period-gap. The eclipse was also confirmed by \citet{Rivera2017} in \textit{HST} data (in the red filter R$_\mathrm{625}$).

This is the first time that the source is detected using \ero{} data. Although \src{} falls within the dense central region of 1.7$^\prime$ that \ero{} cannot spatially resolve, we detected the same eclipse profile from the source events extracted from a 15$^{\prime\prime}$-radius circular region. The folded \ero{} light curves resemble those of \chandra{} (cf. e.g. Fig.\,\ref{profiles} and \ref{fig:eROSITAsignal}), hence confirming the identification of \src{}.

Our analysis did not detect (above the 3.5$\sigma$ threshold) any other periodicity, neither in \chandra{} nor in \ero{} data. The detection of the 3846.15~s modulation of \citet{Bao2023} can be probably attributed to the different approach (peak-removal of the fundamental and first harmonic of the orbital period) adopted by the authors. Our in-depth analysis of the candidate signal identified by \cite{Bao2023} shows that this modulation is unrelated to the source and/or not coherent (see Fig.\,\ref{checkPspin}), indicating that it is not the spin period of the WD. Moreover, we note that the power spectrum shows an excess of power in the frequency range 1--4$\times10^{-4}$~Hz, with several peaks below the detection threshold of 3.5$\sigma$. Epoch-folding at each of these individual peaks lead to all sort of sinusoidal profiles. Due to the lack of detection of other significant periodicities, we classify the source as a candidate polar CV, rather than an IP. 

The source displays a double-humped phase-folded profile, which is also sometimes observed in polar CVs, as seen for example in AM Her \citep{Heise1985} and V496 UMa \citep{Kennedy2022}. This profile shape is typically attributed to the emission from both magnetic poles of the WD, with one being more intense than the other, due to the inclination at which the WD is observed. Concerning \src{}, the energy-resolved folded light curves of the 2002 data sets (e.g. Fig.~\ref{profiles_energyranges}) show similar amplitudes of the two peaks at hard energies, while they are unequal at soft energies. This would suggest that the less active pole is harder than the main pole. The hardness ratio of the source does not give further information, being constant within the error bars. The high cycle-to-cycle variability (Figs.~\ref{profiles}), despite no significant changes in the X-ray flux  (Fig.~\ref{fig:lc_total}), would point to a behaviour typical of magnetic CVs of the polar type, as observed for instance for the eclipsing polars 2PBC J0658.0-1746 \citep{Bernardini2019_polar} or 3XMM J00511.8+634018 \citep{Schwope+20}, which also fall in the period gap.

\subsection{The CV populations in GCs}

47 Tuc has the highest number of CVs among all Galactic GCs, with a total of 43 CVs and candidate CVs identified so far in \textit{HST} optical and NUV data \citep{Rivera2017}. Among them, several have also been detected and identified in X-rays, thanks to \chandra{} and \textit{eROSITA} data \citep{Edmonds2003a,Edmonds2003b,Heinke2005,Bao2023, Saeedi2022}. For instance, the most recent work claimed 11 X-ray CVs in the cluster, identifying the CVs based on their periodic signals \citep{Bao2023}.

We confirm that \src{} has the second shortest orbital period ($\sim$2.4~hr) of all the X-ray CVs in the cluster. 4 out of the 11 CVs have periods within the period gap, close to the ratio for the CVs in the Galaxy bulge, but higher than that for CVs in the solar neighbourhood \citep[20\% and 8\%, respectively, as reported by ][]{Bao2023}. However, it should be noted that the period gap for magnetic CVs is much less marked than that of non-magnetic systems \citep{Webbink&Wickramasinghe02}. Remarkably, as noted by \citet{Bao2023}, in 47 Tuc all the CVs in the period gap are located within the half-light radius (3.17$^\prime$) and the core radius (0.36$^\prime$) of the cluster, while longer-period CVs are more centrally concentrated and are found within the core radius. Moreover, no known CV in the GC has a period below the period gap, and the majority of them show longer periods than those of the CVs in the Galaxy bulge and solar neighbourhood. These peculiarities would suggest different formation channels for the GC CVs compared to those in the Galactic bulge and the solar neighbourhood, contributing to the idea that dynamical encounters may have played a significant role in the GC history \citep[e.g.][]{Belloni2019}. This might be especially true for those CVs outside the core radius, such as \src{}, although a selection bias cannot be entirely excluded (CVs below the period gap are generally less bright). 
However, observational biases might be present towards the brightest X-ray CVs (i.e. magnetic CVs), which appear to be more abundant in GCs than non-magnetic CVs. Whether this is true or not is still a matter of debate \citep{Knigge2012, BelloniRivera2021}. 

Apart from 47 Tuc, three other GCs have a high number of hosted CVs: NGC 6397, NGC 6752, and $\omega$ Cen \citep[see][for a review]{BelloniRivera2021}. All these GCs show a bi-modal distribution of the faintest and the brightest CVs when observed in X-rays. However, while in core-collapsed GCs this distribution persists also in optical, in non-core-collapsed GCs, like 47 Tuc, there is no evidence of bimodality at optical wavelengths. Another difference between these GCs lies in the spatial distribution of their CVs. In NGC 6397 and NGC 6752 the bright CVs are concentrated more towards the centre than faint CVs. The same holds for $\omega$ Cen, though less evident. Instead, in 47 Tuc CVs are uniformly distributed. This difference can be due to the evolution and relaxation time of the cluster.
Nonetheless, with a $L_\mathrm{X}\sim10^{32}$\,\lum, \src{} is among the brightest CVs in all the aforementioned GCs. This is consistent with its magnetic nature, as magnetic CVs are typically brighter \citep[e.g.][]{Mukai2017}.

\section{Conclusions}
\label{sec:conclusions}

We investigated the nature of the CV \src{} in the Galactic GC 47 Tuc. The source shows characteristics which are common among magnetic CVs: a luminosity of $\sim10^{32}$ \lum{}, a temperature of 10~keV, and an orbital period of 8649~s (2.4~hr). The flux of the source is found to be constant in almost all \chandra{} observations, taken in 2000, 2002, 2014--15, and 2022. In the latest 2022 observation, \src{} is not detected, but the upper limit on its flux is consistent with the previous observation in the same epoch. 
Our search for pulsations confirms the previous orbital period and finds no other significant signal. The source is also detected for the first time in \ero{} data and the same orbital period is recovered. This source has been proposed to have a second, shorter period of 3846\,s, indicating that it might be an asynchronous magnetic CV of the IP type. However, our analysis does not confirm the presence of this second period, based also on the low coherence of the candidate signal which is inconsistent with being originated by a spin modulation. Instead, the cycle-to-cycle variability of the amplitude at the 2.4 h period, as well as its X-ray luminosity, suggests that it is a polar.

\begin{acknowledgements}
This research is based on data obtained from the \chandra\ Data Archive and has made use of the software package CIAO provided by the \chandra\ X-ray Center (CXC). The \ero\ data shown here were processed using the eSASS software system developed by the German \ero\ consortium. The authors thank the anonymous referee whose constructive feedback significantly improved the scientific quality and the readability of this manuscript. RA and GLI acknowledge financial support from INAF through grant ``INAF-Astronomy Fellowships in Italy 2022 - (GOG)''. MI is supported by the AASS Ph.D. joint research programme between the University of Rome "Sapienza" and the University of Rome "Tor Vergata", with the collaboration of the National Institute of Astrophysics (INAF). PE and GLI acknowledge financial support from the Italian Ministry for University and Research, through the grants 2017LJ39LM (UNIAM) and 2022Y2T94C (SEAWIND), and from INAF through LG 2023 BLOSSOM. DdM acknowledges support from INAF Astrofund2022. RA and NW are grateful for support provided by the CNES for this work.
\end{acknowledgements}

\bibliographystyle{aa} % style aa.bst
\bibliography{biblio} % your references Yourfile.bib

\begin{thebibliography}{43}
\expandafter\ifx\csname natexlab\endcsname\relax\def\natexlab#1{#1}\fi

\bibitem[{{Arnaud}(1996)}]{Arnaud1996}
{Arnaud}, K.~A. 1996, in Astronomical Society of the Pacific Conference Series,
  Vol. 101, Astronomical Data Analysis Software and Systems V, ed. G.~H.
  {Jacoby} \& J.~{Barnes}, 17

\bibitem[{{Bachetti}(2018)}]{Bachetti2018}
{Bachetti}, M. 2018, {HENDRICS: High ENergy Data Reduction Interface from the
  Command Shell}, Astrophysics Source Code Library, record ascl:1805.019

\bibitem[{{Bao} {et~al.}(2023){Bao}, {Li}, \& {Cheng}}]{Bao2023}
{Bao}, T., {Li}, Z., \& {Cheng}, Z. 2023, \mnras, 521, 4257

\bibitem[{{Baumgardt} \& {Vasiliev}(2021)}]{BaumgardtVasiliev2021}
{Baumgardt}, H. \& {Vasiliev}, E. 2021, \mnras, 505, 5957

\bibitem[{{Belloni} {et~al.}(2019){Belloni}, {Giersz}, {Rivera Sandoval},
  {Askar}, \& {Cieciel{\r{a}}g}}]{Belloni2019}
{Belloni}, D., {Giersz}, M., {Rivera Sandoval}, L.~E., {Askar}, A., \&
  {Cieciel{\r{a}}g}, P. 2019, \mnras, 483, 315

\bibitem[{{Belloni} \& {Rivera}(2021)}]{BelloniRivera2021}
{Belloni}, D. \& {Rivera}, L. 2021, in The Golden Age of Cataclysmic Variables
  and Related Objects V, Vol. 2-7, 13

\bibitem[{{Bernardini} {et~al.}(2019){Bernardini}, {de Martino}, {Mukai},
  {Falanga}, \& {Masetti}}]{Bernardini2019_polar}
{Bernardini}, F., {de Martino}, D., {Mukai}, K., {Falanga}, M., \& {Masetti},
  N. 2019, \mnras, 489, 1044

\bibitem[{{Bhattacharya} {et~al.}(2017){Bhattacharya}, {Heinke}, {Chugunov},
  {Freire}, {Ridolfi}, \& {Bogdanov}}]{Bhattacharya2017}
{Bhattacharya}, S., {Heinke}, C.~O., {Chugunov}, A.~I., {et~al.} 2017, \mnras,
  472, 3706

\bibitem[{{Brunner} {et~al.}(2022){Brunner}, {Liu}, {Lamer}, {Georgakakis},
  {Merloni}, {Brusa}, {Bulbul}, {Dennerl}, {Friedrich}, {Liu}, {Maitra},
  {Nandra}, {Ramos-Ceja}, {Sanders}, {Stewart}, {Boller}, {Buchner}, {Clerc},
  {Comparat}, {Dwelly}, {Eckert}, {Finoguenov}, {Freyberg}, {Ghirardini},
  {Gueguen}, {Haberl}, {Kreykenbohm}, {Krumpe}, {Osterhage}, {Pacaud},
  {Predehl}, {Reiprich}, {Robrade}, {Salvato}, {Santangelo}, {Schrabback},
  {Schwope}, \& {Wilms}}]{Brunner2022}
{Brunner}, H., {Liu}, T., {Lamer}, G., {et~al.} 2022, A\&A, 661, A1

\bibitem[{{Buccheri} {et~al.}(1983){Buccheri}, {Bennett}, {Bignami}, {Bloemen},
  {Boriakoff}, {Caraveo}, {Hermsen}, {Kanbach}, {Manchester}, {Masnou},
  {Mayer-Hasselwander}, {{\"O}zel}, {Paul}, {Sacco}, {Scarsi}, \&
  {Strong}}]{Buccheri1983}
{Buccheri}, R., {Bennett}, K., {Bignami}, G.~F., {et~al.} 1983, A\&A, 128, 245

\bibitem[{{Chen} {et~al.}(2018){Chen}, {Richer}, {Caiazzo}, \&
  {Heyl}}]{Chen2018}
{Chen}, S., {Richer}, H., {Caiazzo}, I., \& {Heyl}, J. 2018, \apj, 867, 132

\bibitem[{{Edmonds} {et~al.}(2003{\natexlab{a}}){Edmonds}, {Gilliland},
  {Heinke}, \& {Grindlay}}]{Edmonds2003a}
{Edmonds}, P.~D., {Gilliland}, R.~L., {Heinke}, C.~O., \& {Grindlay}, J.~E.
  2003{\natexlab{a}}, Astrophysical Journal, 596, 1177

\bibitem[{{Edmonds} {et~al.}(2003{\natexlab{b}}){Edmonds}, {Gilliland},
  {Heinke}, \& {Grindlay}}]{Edmonds2003b}
{Edmonds}, P.~D., {Gilliland}, R.~L., {Heinke}, C.~O., \& {Grindlay}, J.~E.
  2003{\natexlab{b}}, Astrophysical Journal, 596, 1197

\bibitem[{{Garc{\'\i}a-Berro} {et~al.}(2014){Garc{\'\i}a-Berro}, {Torres},
  {Althaus}, \& {Miller Bertolami}}]{GarciaBerro2014}
{Garc{\'\i}a-Berro}, E., {Torres}, S., {Althaus}, L.~G., \& {Miller Bertolami},
  M.~M. 2014, \aap, 571, A56

\bibitem[{Gregory \& Loredo(1992)}]{Gregory1992}
Gregory, P.~C. \& Loredo, T.~J. 1992, ApJ, 398, 146

\bibitem[{{Grindlay} {et~al.}(2001){Grindlay}, {Heinke}, {Edmonds}, \&
  {Murray}}]{Grindlay2001}
{Grindlay}, J.~E., {Heinke}, C., {Edmonds}, P.~D., \& {Murray}, S.~S. 2001,
  Science, 292, 2290

\bibitem[{{Heinke} {et~al.}(2005){Heinke}, {Grindlay}, {Edmonds}, {Cohn},
  {Lugger}, {Camilo}, {Bogdanov}, \& {Freire}}]{Heinke2005}
{Heinke}, C.~O., {Grindlay}, J.~E., {Edmonds}, P.~D., {et~al.} 2005,
  Astrophysical Journal, 625, 796

\bibitem[{{Heise} {et~al.}(1985){Heise}, {Brinkman}, {Gronenschild}, {Watson},
  {King}, {Stella}, \& {Kieboom}}]{Heise1985}
{Heise}, J., {Brinkman}, A.~C., {Gronenschild}, E., {et~al.} 1985, \aap, 148,
  L14

\bibitem[{{Israel} {et~al.}(2016){Israel}, {Esposito}, {Rodr{\'{\i}}guez
  Castillo}, \& {Sidoli}}]{Israel2016}
{Israel}, G.~L., {Esposito}, P., {Rodr{\'{\i}}guez Castillo}, G.~A., \&
  {Sidoli}, L. 2016, MNRAS, 462, 4371

\bibitem[{Israel \& Stella(1996)}]{Israel1996}
Israel, G.~L. \& Stella, L. 1996, ApJ, 468, 369

\bibitem[{{Kennedy} {et~al.}(2022){Kennedy}, {Littlefield}, \&
  {Garnavich}}]{Kennedy2022}
{Kennedy}, M.~R., {Littlefield}, C., \& {Garnavich}, P.~M. 2022, \mnras, 513,
  2930

\bibitem[{{Knigge}(2012)}]{Knigge2012}
{Knigge}, C. 2012, \memsai, 83, 549

\bibitem[{{Lugaro} {et~al.}(2013){Lugaro}, {D'Orazi}, {Campbell}, {Doherty},
  {Lattanzio}, {Pignatari}, \& {Carretta}}]{Lugaro+2013}
{Lugaro}, M., {D'Orazi}, V., {Campbell}, S.~W., {et~al.} 2013, \memsai, 84, 109

\bibitem[{{Marks} \& {Kroupa}(2010)}]{Marks2010}
{Marks}, M. \& {Kroupa}, P. 2010, \mnras, 406, 2000

\bibitem[{{Mewe}(1991)}]{Mewe1991}
{Mewe}, R. 1991, \aapr, 3, 127

\bibitem[{{Meylan} \& {Heggie}(1997)}]{MeylanHeggie1997}
{Meylan}, G. \& {Heggie}, D.~C. 1997, \aapr, 8, 1

\bibitem[{{Mukai}(2017)}]{Mukai2017}
{Mukai}, K. 2017, Publications of the Astronomical Society of the Pacific, 129,
  062001

\bibitem[{{Pooley} \& {Hut}(2006)}]{PooleyHut2006}
{Pooley}, D. \& {Hut}, P. 2006, \apjl, 646, L143

\bibitem[{{Pooley} {et~al.}(2003){Pooley}, {Lewin}, {Anderson}, {Baumgardt},
  {Filippenko}, {Gaensler}, {Homer}, {Hut}, {Kaspi}, {Makino}, {Margon},
  {McMillan}, {Portegies Zwart}, {van der Klis}, \& {Verbunt}}]{Pooley2003}
{Pooley}, D., {Lewin}, W. H.~G., {Anderson}, S.~F., {et~al.} 2003, \apjl, 591,
  L131

\bibitem[{{Predehl} {et~al.}(2021){Predehl}, {Andritschke}, {Arefiev},
  {Babyshkin}, {Batanov}, {Becker}, {B{\"o}hringer}, {Bogomolov}, {Boller},
  {Borm}, {Bornemann}, {Br{\"a}uninger}, {Br{\"u}ggen}, {Brunner}, {Brusa},
  {Bulbul}, {Buntov}, {Burwitz}, {Burkert}, {Clerc}, {Churazov}, {Coutinho},
  {Dauser}, {Dennerl}, {Doroshenko}, {Eder}, {Emberger}, {Eraerds},
  {Finoguenov}, {Freyberg}, {Friedrich}, {Friedrich}, {F{\"u}rmetz},
  {Georgakakis}, {Gilfanov}, {Granato}, {Grossberger}, {Gueguen}, {Gureev},
  {Haberl}, {H{\"a}lker}, {Hartner}, {Hasinger}, {Huber}, {Ji}, {Kienlin},
  {Kink}, {Korotkov}, {Kreykenbohm}, {Lamer}, {Lomakin}, {Lapshov}, {Liu},
  {Maitra}, {Meidinger}, {Menz}, {Merloni}, {Mernik}, {Mican}, {Mohr},
  {M{\"u}ller}, {Nandra}, {Nazarov}, {Pacaud}, {Pavlinsky}, {Perinati},
  {Pfeffermann}, {Pietschner}, {Ramos-Ceja}, {Rau}, {Reiffers}, {Reiprich},
  {Robrade}, {Salvato}, {Sanders}, {Santangelo}, {Sasaki}, {Scheuerle},
  {Schmid}, {Schmitt}, {Schwope}, {Shirshakov}, {Steinmetz}, {Stewart},
  {Str{\"u}der}, {Sunyaev}, {Tenzer}, {Tiedemann}, {Tr{\"u}mper}, {Voron},
  {Weber}, {Wilms}, \& {Yaroshenko}}]{Predehl2021}
{Predehl}, P., {Andritschke}, R., {Arefiev}, V., {et~al.} 2021, \aap, 647, A1

\bibitem[{{Rivera Sandoval} {et~al.}(2018){Rivera Sandoval}, {van den Berg},
  {Heinke}, {Cohn}, {Lugger}, {Anderson}, {Cool}, {Edmonds}, {Wijnands},
  {Ivanova}, \& {Grindlay}}]{Rivera2017}
{Rivera Sandoval}, L.~E., {van den Berg}, M., {Heinke}, C.~O., {et~al.} 2018,
  \mnras, 475, 4841

\bibitem[{{Saeedi} {et~al.}(2022){Saeedi}, {Liu}, {Knies}, {Sasaki}, {Becker},
  {Bulbul}, {Dennerl}, {Freyberg}, {Laktionov}, \& {Merloni}}]{Saeedi2022}
{Saeedi}, S., {Liu}, T., {Knies}, J., {et~al.} 2022, \aap, 661, A35

\bibitem[{{Schwope} {et~al.}(2020){Schwope}, {Worpel}, {Webb}, {Koliopanos}, \&
  {Guillot}}]{Schwope+20}
{Schwope}, A.~D., {Worpel}, H., {Webb}, N.~A., {Koliopanos}, F., \& {Guillot},
  S. 2020, \aap, 637, A35

\bibitem[{{Simunovic} {et~al.}(2023){Simunovic}, {Puzia}, {Miller}, {Carrasco},
  {Dotter}, {Cassisi}, {Monty}, \& {Stetson}}]{Simunovic2023}
{Simunovic}, M., {Puzia}, T.~H., {Miller}, B., {et~al.} 2023, arXiv e-prints,
  arXiv:2304.11240

\bibitem[{{Strader} {et~al.}(2012){Strader}, {Chomiuk}, {Maccarone},
  {Miller-Jones}, {Seth}, {Heinke}, \& {Sivakoff}}]{Strader+2012}
{Strader}, J., {Chomiuk}, L., {Maccarone}, T.~J., {et~al.} 2012, \apjl, 750,
  L27

\bibitem[{{Thompson} {et~al.}(2020){Thompson}, {Udalski}, {Dotter}, {Rozyczka},
  {Schwarzenberg-Czerny}, {Pych}, {Beletsky}, {Burley}, {Marshall},
  {McWilliam}, {Morrell}, {Osip}, {Monson}, {Persson}, {Szyma{\'n}ski},
  {Soszy{\'n}ski}, {Poleski}, {Ulaczyk}, {Wyrzykowski}, {Koz{\l}owski},
  {Mr{\'o}z}, {Pietrukowicz}, \& {Skowron}}]{Thompson+2020}
{Thompson}, I.~B., {Udalski}, A., {Dotter}, A., {et~al.} 2020, \mnras, 492,
  4254

\bibitem[{{Thygesen} {et~al.}(2014){Thygesen}, {Sbordone}, {Andrievsky},
  {Korotin}, {Yong}, {Zaggia}, {Ludwig}, {Collet}, {Asplund}, {Ventura},
  {D'Antona}, {Mel{\'e}ndez}, \& {D'Ercole}}]{Thygesen2014}
{Thygesen}, A.~O., {Sbordone}, L., {Andrievsky}, S., {et~al.} 2014, \aap, 572,
  A108

\bibitem[{{Truemper}(1982)}]{Truemper1982}
{Truemper}, J. 1982, Advances in Space Research, 2, 241

\bibitem[{{Verner} {et~al.}(1996){Verner}, {Ferland}, {Korista}, \&
  {Yakovlev}}]{Verner1996}
{Verner}, D.~A., {Ferland}, G.~J., {Korista}, K.~T., \& {Yakovlev}, D.~G. 1996,
  \apj, 465, 487

\bibitem[{{Webbink} \& {Wickramasinghe}(2002)}]{Webbink&Wickramasinghe02}
{Webbink}, R.~F. \& {Wickramasinghe}, D.~T. 2002, \mnras, 335, 1

\bibitem[{{Weisskopf} {et~al.}(2000){Weisskopf}, {Tananbaum}, {Van Speybroeck},
  \& {O'Dell}}]{Weisskopf2000}
{Weisskopf}, M.~C., {Tananbaum}, H.~D., {Van Speybroeck}, L.~P., \& {O'Dell},
  S.~L. 2000, in Society of Photo-Optical Instrumentation Engineers (SPIE)
  Conference Series, Vol. 4012, X-Ray Optics, Instruments, and Missions III,
  ed. J.~E. {Truemper} \& B.~{Aschenbach}, 2--16

\bibitem[{{Wilms} {et~al.}(2000){Wilms}, {Allen}, \& {McCray}}]{Wilms2000}
{Wilms}, J., {Allen}, A., \& {McCray}, R. 2000, \apj, 542, 914

\bibitem[{{Zoccali} {et~al.}(2001){Zoccali}, {Renzini}, {Ortolani},
  {Bragaglia}, {Bohlin}, {Carretta}, {Ferraro}, {Gilmozzi}, {Holberg},
  {Marconi}, {Rich}, \& {Wesemael}}]{Zoccali+2001}
{Zoccali}, M., {Renzini}, A., {Ortolani}, S., {et~al.} 2001, \apj, 553, 733

\end{thebibliography}

\begin{appendix}

\section{\chandra{} observations of 47 Tuc}
\label{sec:appendixA}

\begin{table*}[h]
\caption{Log of \textit{Chandra} observations in chronological order. Upper limit on the latest observation is given at 90\%c.l.}
\label{tab:logChandra}
\centering
%\begin{small}
    \begin{tabular}{lccccc}
    \hline\hline\noalign{\smallskip}
    Obs. ID & Date & Exp. (ks) & Net count rate ($10^{-3}$ cts/s) & Instrument & Ref.\\ %qua si potrebbero aggiungere anche i pixel utilizzati
    \noalign{\smallskip}\hline\noalign{\smallskip}
    78 & 2000-03-16 07:17:27 & 3.88 & $5.9\pm1.2$ & ACIS-I & (a)\\
    \noalign{\smallskip}
    953 & 2000-03-16 08:38:40 & 31.68 & $1.2\pm0.2$ & ACIS-I & (a,b)\\
    \noalign{\smallskip}
    954 & 2000-03-16 18:01:59 & 0.85 & -- & ACIS-I & (a)\\ 
    \noalign{\smallskip}
    955 & 2000-03-16 18:32:00 & 31.68 & $4.5\pm0.4$ & ACIS-I & (a,b)\\
    \noalign{\smallskip}
    956 & 2000-03-17 03:55:20 & 4.69 & $5.3\pm1.1$ & ACIS-I & (a)\\
    \noalign{\smallskip}
    2735 & 2002-09-29 16:57:56 & 65.24 & $6.4\pm0.3$ & ACIS-S & (a,b)\\
    \noalign{\smallskip}
    3384 & 2002-09-30 11:37:18 & 5.31 & $9.0\pm1.3$ & ACIS-S & (a)\\
    \noalign{\smallskip}
    2736 & 2002-09-30 13:24:28 & 65.24 & $6.0\pm0.3$ & ACIS-S & (a,b)\\
    \noalign{\smallskip}
    3385 & 2002-10-01 08:12:28 & 5.31 & $8.8\pm1.4$ & ACIS-S & (a)\\
    \noalign{\smallskip}
    2737 & 2002-10-02 18:50:07 & 65.24 & $7.6\pm0.3$ & ACIS-S & (a,b)\\
    \noalign{\smallskip}
    3386 & 2002-10-03 13:37:18 & 5.55 & $2.1\pm0.6$ & ACIS-S & (a)\\
    \noalign{\smallskip}
    2738 & 2002-10-11 01:41:55 & 68.77 & $7.4\pm0.3$ & ACIS-S & (a,b)\\
    \noalign{\smallskip}
    3387 & 2002-10-11 21:22:09 & 5.74 & $2.1\pm0.6$ & ACIS-S & (a)\\
    \noalign{\smallskip}
    16527 & 2014-09-05 04:38:37 & 40.88 & $3.1\pm0.3$ & ACIS-S\\
    \noalign{\smallskip}
    15747 & 2014-09-09 19:32:57 & 50.04 & $4.4\pm0.3$ & ACIS-S\\
    \noalign{\smallskip}
    16529 & 2014-09-21 07:55:51 & 24.7 & $3.9\pm0.4$ & ACIS-S\\
    \noalign{\smallskip}
    17420 & 2014-09-30 22:56:03 & 9.13 & $2.7\pm0.6$ & ACIS-S\\
    \noalign{\smallskip}
    15748 & 2014-10-02 06:17:00 & 16.24 & $4.5\pm0.5$ & ACIS-S\\
    \noalign{\smallskip}
    16528 & 2015-02-02 14:23:34 & 40.28 & -- & ACIS-S\\ 
    \noalign{\smallskip}
    26229 & 2022-01-26 15:20:22 & 9.65 & $1.0\pm0.4$ & ACIS-S\\
    \noalign{\smallskip}
    26286 & 2022-01-27 02:10:56 & 9.83 & $\leq$1.2 & ACIS-S\\	
    \noalign{\smallskip}\hline
    \end{tabular}
%\end{small}
\tablefoot{(a) \citet{Heinke2005}, (b) \citet{Israel2016}.
   }
\end{table*}

\clearpage

\section{Best-fit results}
\label{sec:appendixB}

The following tables show the best-fit parameters obtained from fitting the seven data sets 955, 2735, 2736, 2737, 2738, 15747, 16527 individually (columns 2--6) and from imposing the same local absorption $N_\mathrm{H}^\mathrm{47 Tuc + local}$ (columns 7--8). 

\begin{table*}
\caption{Best-fit values of the seven data sets 955, 2735, 2736, 2737, 2738, 15747, 16527 with the power law model, assuming $N_\mathrm{H}^\mathrm{GAL} = 5.5 \times 10^{20}$ cm$^{-2}$, in two different cases: independent fits for each spectrum (columns 2--6) and simultaneous fit with the same local absorption (columns 7--8). The latter fit resulted in $N_\mathrm{H}^\mathrm{47 Tuc + local}=(2.6^{+1.2}_{-1.1})\times 10^{21}$ cm$^{-2}$, with $\chi^2_{red}$(d.o.f.)$=$1.22(81) and n.h.p.$=$0.08. Fluxes are computed in the energy range 0.5--6\,keV.}
\centering
\begin{tabular}{rccccccc}
\hline\hline\noalign{\smallskip}
(1) & (2)   & (3)   & (4)   & (5)   & (6)   & (7)   & (8)   \\
Obs. &  $N_\mathrm{H}^\mathrm{47 Tuc + local}$ & $\Gamma$ & $F_\mathrm{abs}$ & $\chi^2_{red}$(d.o.f.)\tablefootmark{a} & n.h.p.\tablefootmark{b} & $\Gamma$ & $F_\mathrm{abs}$\\ 
\noalign{\smallskip} & ($10^{21}$ cm$^{-2}$) & & ($10^{-14}$ \flux{}) &&&& ($10^{-14}$ \flux{})\\
\noalign{\smallskip}\cmidrule(lr){1-1}\cmidrule(lr){2-6}\cmidrule(lr){7-8}\noalign{\smallskip}
955 & 2.0$^{+14.2}_{-2.0}$ & 1.3$^{+1.1}_{-0.5}$ & 8 $\pm$ 4 & 0.18(2) & 0.84 & 1.3$^{+0.5}_{-0.4}$ & 7$^{+4}_{-2}$ \\  \noalign{\smallskip}
2735 & 1.7$^{+2.2}_{-1.7}$ & 1.4$^{+0.3}_{-0.2}$ & 7 $\pm $1 & 1.03(15) & 0.42 & 1.5 $\pm$ 0.2 & 6.4$^{+1.2}_{-0.8}$ \\   \noalign{\smallskip}
2736 & 4.0$^{+3.7}_{-2.9}$ & 1.6 $\pm$ 0.3 & 6 $\pm$ 1 & 1.31(14) & 0.19 & 1.5 $\pm$ 0.2 & 6.2$^{+1.0}_{-0.8}$ \\	\noalign{\smallskip}
2737 & 3.8$^{+2.6}_{-2.2}$ & 1.5$^{+0.3}_{-0.2}$ & 8 $\pm$ 1 & 2.02(18) & 0.01 & 1.4 $\pm$ 0.2 & 8 $\pm$ 1 \\ \noalign{\smallskip}
2738 & 2.7$^{+2.5}_{-2.3}$ & 1.8 $\pm$ 0.3 & 7 $\pm$ 1 & 0.93(18) & 0.54 & 1.75 $\pm$ 0.2 & 7.0$^{+1.2}_{-0.8}$ \\    \noalign{\smallskip}
15747 & 2.9 (frozen) & 1.7 $\pm$ 0.3 & 5 $\pm$ 1 & 0.47(7) & 0.86 & 1.65 $\pm$ 0.3 & 5.1$^{+1.1}_{-0.8}$ \\	\noalign{\smallskip}
16527 & 2.9 (frozen) & 1.3$^{+0.5}_{-0.4}$ & 4$^{+2}_{-1}$ & 2.33(3) & 0.07 & 1.3$^{+0.5}_{-0.4}$ & 4$^{+2}_{-1}$ \\	\noalign{\smallskip}
%16529 & 0.2 (frozen) & 3 $\pm$ 1 & 4 $\pm$ 1 & 0.06(1) & 0.81\\
\noalign{\smallskip}\hline\hline
\end{tabular}
\tablefoot{\tablefoottext{a}{Degrees of freedom (d.o.f).} \tablefoottext{b}{Null hypothesis probability (n.h.p.).}
}
\label{tab:FitPwrlaw}
\end{table*}

\begin{table*}
\centering
\caption{As before, for the thermal Bremsstrahlung model. For the simultaneous fit with the same local absorption (columns 7--8) the best-fit local absorption was of $N_\mathrm{H}^\mathrm{47 Tuc + local} = (2.0\pm 0.9)\times 10^{21}$ cm$^{-2}$,  with $\chi^2_{red}$(d.o.f.)$=$1.21(81) and n.h.p.$=$0.1.}
%\begin{footnotesize}
\begin{tabular}{rccccccc}
\hline\hline\noalign{\smallskip}
(1) & (2)   & (3)   & (4)   & (5)   & (6)   & (7)   & (8)   \\
Obs. & $N_\mathrm{H}^\mathrm{47 Tuc + local}$ & kT & $F_\mathrm{abs}$ & $\chi^2_{red}$(d.o.f.) & n.h.p. & kT & $F_\mathrm{abs}$ \\
\noalign{\smallskip} & ($10^{21}$ cm$^{-2}$) & (keV) & ($10^{-14}$ \flux{}) &&& (keV) & [$10^{-14}$ \flux{}]\\
\noalign{\smallskip}\cmidrule(lr){1-1}\cmidrule(lr){2-6}\cmidrule(lr){7-8}\noalign{\smallskip}
955 & 1.7$^{+11.3}_{-1.7}$ & 60$^{+50}_{-30}$ & 7$^{+16}_{-7}$ & 0.17(2) & 0.84 & 46$^{+45}_{-24}$ & 7$^{+3}_{-7}$ \\ \noalign{\smallskip}
2735 & 1.3$^{+1.8}_{-1.3}$ & 18$^{+66}_{-11}$ & 6$^{+1}_{-6}$ & 1.01(15) & 0.44 & 13$_{-7}^{+35}$ & 6$^{+1}_{-2}$ \\	\noalign{\smallskip}
2736 & 3.3$_{-2.3}^{+2.9}$ & 9$_{-5}^{+26}$ & 5.4$^{+0.6}_{-2.4}$ & 1.22(14) & 0.25 & 13$_{-6}^{+37}$ & 5.6$^{+0.8}_{-2.6}$ \\	\noalign{\smallskip}
2737 & 3.1$^{+2.1}_{-1.7}$ & 12$_{-6}^{+54}$ & 7$^{+1}_{-5}$ & 1.99(18) & 0.01 & 20$_{-10}^{+140}$ & 8$^{+1}_{-6}$ \\	\noalign{\smallskip}
2738 & 1.5$_{-1.5}^{+2.0}$ & 6$_{-3}^{+7}$ & 6.3$^{+0.8}_{-2.0}$ & 0.92(18) & 0.55 & 5$_{-2}^{+4}$ & 6$^{+1}_{-2}$ \\ \noalign{\smallskip}
15747 & 2.2 (frozen) & 8$_{-4}^{+30}$ & 4.6$^{+0.7}_{-2.4}$ & 0.57(7) & 0.78 & 8$_{-4}^{+37}$ & 4.7$^{+0.8}_{-1.5}$ \\  \noalign{\smallskip}
16527 & 2.2 (frozen) & 38$^{+44}_{-23}$ & 4$^{+1}_{-4}$ & 2.30(3) & 0.07 & 42$^{+47}_{-25}$ & 4$^{+2}_{-4}$ \\    \noalign{\smallskip}\hline\hline
\end{tabular}
\label{tab:FitBrems}
\end{table*}

\begin{table*}
\centering
\caption{As before, for the \texttt{vmekal} model. The simultaneous fit of the seven spectra with the same local absorption (columns 7--8) resulted in $N_\mathrm{H}^\mathrm{47 Tuc + local} = (1.9 \pm 0.9) \times 10^{21}$ cm$^{-2}$,  with $\chi^2_{red}$(d.o.f.)$ = $1.22(81) and n.h.p.$=$0.09.}
\begin{tabular}{rccccccc}
\hline\hline\noalign{\smallskip}
(1) & (2)   & (3)   & (4)   & (5)   & (6)   & (7)   & (8)   \\
Obs. & $N_\mathrm{H}^\mathrm{47 Tuc + local}$ & kT & $F_\mathrm{abs}$ & $\chi^2_{red}$(d.o.f.) & n.h.p. & kT & $F_\mathrm{abs}$ \\
\noalign{\smallskip} & ($10^{21}$ cm$^{-2}$) & (keV) & ($10^{-14}$ \flux{}) &&& (keV) & ($10^{-14}$ \flux{})\\
\noalign{\smallskip}\cmidrule(lr){1-1}\cmidrule(lr){2-6}\cmidrule(lr){7-8}\noalign{\smallskip}
955 & $1.8^{+12.5}_{-1.8}$ & 50$\pm$30 & $7^{+13}_{-7}$ & 0.17(2) & 0.84 & 50$^{+30}_{-46}$ & 7$^{+9}_{-7}$ \\ \noalign{\smallskip}
2735 & $1.2_{-1.2}^{+1.8}$ & 19$^{+61}_{-12}$ & $6.5_{-6.0}^{+0.9}$ & 1.01(15) & 0.44 & 13$_{-6}^{+36}$ & 6$_{-5}^{+1}$ \\	\noalign{\smallskip}
2736 & $3.3_{-2.3}^{+2.8}$ & $9_{-4}^{+23}$ & $5.4_{-2.1}^{+0.8}$ & 1.21(14) & 0.26 & 13$_{-6}^{+35}$ & 5.7$_{-3.0}^{+0.8}$ \\ \noalign{\smallskip}
2737 & 3.1$^{+2.0}_{-1.8}$ & $12_{-6}^{+55}$ & $7.2_{-2.7}^{+0.8}$ & 2.01(18) & 0.01 & 19$^{+61}_{-10}$ & 8$_{-4}^{+1}$ \\ \noalign{\smallskip}
2738 & $1.4_{-1.4}^{+1.8}$ & $6_{-2}^{+7}$ & $6.4_{-1.6}^{+0.7}$ & 0.93(18) & 0.55 & 5$_{-1}^{+4}$ & 6.2$_{-1.2}^{+0.7}$ \\ \noalign{\smallskip}
15747 & 2.2 (frozen) & $8_{-4}^{+30}$ & $4.7_{-2.3}^{+0.8}$ & 0.57(7) & 0.78 & 8$_{-4}^{+36}$ & 4.8$_{-2.9}^{+0.8}$ \\ \noalign{\smallskip}
16527 & 2.2 (frozen) & 31$^{+49}_{-26}$ & $4^{+3}_{-4}$ & 2.31(3) & 0.07 & 50$\pm$30 & 4$_{-4}^{+5}$ \\ \noalign{\smallskip}\hline\hline
\end{tabular}
\label{tab:FitVmekal}
\end{table*}

\end{appendix}

\end{document}